\documentclass{LMCS}
\usepackage{defns, amssymb, amsmath, latexsym, mathrsfs}
\usepackage{hyperref}

\usepackage{amsthm}

\newtheorem{default}{}[section]

\theoremstyle{plain}\newtheorem{claim}[default]{Claim}
\theoremstyle{plain}\newtheorem{corollary}[default]{Corollary}
\theoremstyle{plain}\newtheorem{definition}[default]{Definition}
\theoremstyle{plain}
\theoremstyle{plain}
\theoremstyle{plain}\newtheorem{exer-hard}[default]{*Exercise}
\theoremstyle{plain}\newtheorem{lemma}[default]{Lemma}
\theoremstyle{plain}
\theoremstyle{plain}
\theoremstyle{plain}
\theoremstyle{plain}\newtheorem{theorem}[default]{Theorem}

\renewcommand{\L}[0]{\ensuremath{\mathbf{L}}}

\def\Remark{{\medskip\noindent{\bf Remark~}}}

\def\qed{{\hfill  $\Box$}}

\def\doi{2 (1:3) 2006}
\lmcsheading%
{\doi}
{40}
{}
{}
{Nov.~30, 2004}
{Mar.~06, 2006}
{}

\begin{document}
\title[Theories for \TCZ\ and other Small Complexity Classes]{Theories for \TCZ\ and other Small Complexity Classes}
\author[Phuong Nguyen]{Phuong Nguyen}
\address{University of Toronto, 10 King's College Road, Toronto,
Ontario M5S 3G4, Canada}
\email{\{ntp,sacook\}@cs.toronto.edu}

\author[Stephen Cook]{Stephen Cook}


\begin{abstract}
We present a general method for introducing finitely axiomatizable
``minimal'' two-sorted theories for various subclasses of \Ptime\ (problems
solvable in polynomial time).  The two sorts are natural
numbers and finite sets of natural numbers.  The latter are essentially
the finite binary strings, which provide a natural domain for defining
the functions and sets in small complexity classes.  We concentrate on the
complexity class \TCZ, whose problems are defined by uniform
polynomial-size families of bounded-depth Boolean circuits with
majority gates.  We present an elegant theory \VTCZ\ in which the
provably-total functions are those associated with \TCZ, and
then prove that \VTCZ\ is ``isomorphic'' to a different-looking single-sorted
theory introduced by Johannsen and Pollet.
The most technical part of the isomorphism proof is defining binary
number multiplication in terms a bit-counting function,
and  showing how to formalize the proofs of its algebraic properties.
\end{abstract}

\maketitle
\vskip-\bigskipamount
\section{Introduction}

Non-uniform \ACZ\ is the class of languages accepted by 
polynomial-size families of
constant-depth Boolean circuits (where the gates have
unbounded fan-in).
Non-uniform \TCZ\ is defined similarly, where the circuits may contain 
{\em majority} gates (i.e., gates with unbounded fan-in, which output 1 if and only
if the number of 1 inputs is more than the number of 0 inputs),
and for non-uniform $\ACZ(m)$%
\footnote{$\ACZ(2)$ is also called $\ACC(2)$ in \cite{Johannsen:98:beam-buss}}
the additional gates are $\mod{m}$ gates,
i.e., gates with unbounded fan-in which output 1
if and only if the number of 1 inputs is exactly 1 modulo $m$.

Each of these classes has a uniform version, where
the families of circuits are uniform.
Here we consider \FO-uniformity \cite{Immerman:99:book},
i.e., each circuit family can be described by some first-order formula.
We will focus on the uniform classes, and will simply use \ACZ, \TCZ\ and $\ACZ(m)$
without the adjective ``uniform''.

Each of these classes can be defined more generally as a class of
{\em relations} rather than languages.
A class {\bf C} is then associated with a {\em function} class {\bf FC}, which
is essentially the set of functions of at most polynomial growth
whose {\em bit graphs} are in {\bf C}.
Then \TCZ\ (resp. \FTCZ) is the class of problems (resp. functions)
$\ACZ$ reducible to the {\em counting function},
which outputs the number of 1 bits in the input string.
The same holds for $\ACZ(m)$ and $\FACZ(m)$, with the {\em modulo $m$} function
instead of the counting function.

It is known that
\begin{equation*}
\ACZ \subsetneq \ACZ(p) \subsetneq \ACZ(pq) \subseteq \ACC \subseteq \TCZ \subseteq \NCOne,
\end{equation*}
for any distinct prime numbers $p,q$, where
$\ACC=\bigcup_{m=2}^\infty \ACZ(m)$.
However it is an open question whether any of last three
inclusions is strict.
It is also unknown, for example, whether $\ACZ(6) \subsetneq \NCOne$,
although $\ACZ(p) \ne \ACZ(q)$ for distinct prime numbers $p, q$.

In this paper we study second-order
logical theories
associated with these and other complexity classes.
We show that our theories \VTCZ\ and $\VZ(m)$ characterize \TCZ\ and $\ACZ(m)$ in the same way
that Buss's theories $\SOneTwo, \STwoTwo, \ldots $ characterize the polynomial
time hierarchy \cite{Buss:86:book}.
Thus we show that \FTCZ\ is precisely the class of \SigOneOne-definable functions
of \VTCZ, and similarly $\FACZ(m)$ is the class of \SigOneOne-definable functions
of $\VZ(m)$.

In Section~\ref{s:RSUV} we show that our theory \VTCZ\ is {\em RSUV isomorphic}
to \DelCR, a ``minimal'' first-order theory that also characterizes \TCZ\
\cite{Johannsen:Pollett:00} but which is defined very differently from
\VTCZ.  Since \VTCZ\ is finitely axiomatizable, it follows that
\DelCR\ is also, and this answers an open question in
\cite{Johannsen:Pollett:00} by showing that there is a constant upper bound
to the nesting depth of the $\DelOneb$ bit-comprehension rule
required to prove theorems in \DelCR.
Our RSUV isomorphism is more difficult than the original ones
given in \cite{Razborov:93:Clote-Krajicek, Takeuti:93:Clote-Krajicek},
as we explain below in Section \ref{s:sec-order-theories}.

The theory \VTCZ\ is obtained by adding to the ``base'' theory \VZ\
\cite{Zambella:96:jsl, Cook:05:quaderni}
(a theory that characterizes \ACZ) the axiom \NUMONES\ which encodes
the counting function which is complete for \TCZ.
This is indeed a generic method that can be used to develop ``minimal'',
finitely axiomatizable theories
characterizing other small classes, including the sequence
$$
  \NCOne\subseteq \L\subseteq \SL\subseteq\NL\subseteq \NC\subseteq\Ptime
$$
In proving that our theories characterize the corresponding classes,
we follow the approach laid down in \cite{Cook:05:quaderni}
which uses ``minimal'', universal theories over the languages of the functions in the
corresponding classes.
The universal counterpart of $\VTCZ$ is called \VTCZbar.
The main tasks remaining are to
(i) show that the universal theories (such as \VTCZbar) are conservative
extensions of the original theories (e.g., \VTCZ); and
(ii) prove the {\em Witnessing Theorems} for the universal theories.
The general results in Theorem~\ref{t:sig0-comp-new-syms}
and Theorem~\ref{t:witnessing} (the General Witnessing Theorem)
should be useful for these purposes in other contexts.

Our universal theories are ``minimal'' theories for the corresponding
complexity classes in the sense that the axioms
consist of straightforward definitions for the functions and predicates
in the class.  For example, $\VTCZbar$ satisfies this condition,
and since it is a conservative extension of \VTCZ, the latter is also
a minimal theory for \TCZ, and so is its first-order counterpart
\DelCR.  However as explained below, the extensions \rbarz and
$\mathbf{C}^0_2$ of \DelCR\ also define precisely the \TCZ\
functions, but they prove $\Sigma^b_2$ theorems which 
(under a cryptographic assumption) are not
provable in \DelCR\ and
hence are apparently not minimal theories for \TCZ.

This paper is based in part on its precursors \cite{Nguyen:04:thesis} and
\cite{Nguyen:Cook:04}.

\subsection{Previous First-Order Theories for our Classes}

In \cite{Clote:Takeuti:95}, Clote and Takeuti introduce the notion of
{\em essentially sharply bounded} (esb) formulas in a theory $\mathcal{T}$.
They introduce the first-order theories \TACZtwo, \TACZsix, and \TTCZ,
and show that a function is esb-definable in one of these theories
iff it is in \ACZ(2), \ACZ(6), or \TCZ, respectively.
However, the notion of esb-definable seems unnecessarily complicated.

In \cite{Johannsen:96:godel}, Johannsen introduces the first-order theory
\rbarz, and shows that
the class of \TCZ\ functions is exactly the class of functions
$\Sigma_1^b$-definable in \rbarz.
In \cite{Johannsen:Pollett:98}, Johannsen and Pollett
introduce a hierarchy $\{\mathbf{C}^0_k\}_{k\ge 1}$ of first-order
theories,
where $\mathbf{C}^0_{k}$ characterizes the class
of functions computable by families of constant-depth threshold circuits of
size bounded by $\tau_k(n)$, where $\tau_1(n) = O(n), \tau_{k+1}(n) = 2^{\tau_k(\log n)}$.
In particular, $\mathbf{C}^0_2$ captures \TCZ.
Later Johannsen and Pollett \cite{Johannsen:Pollett:00} introduce
the ``minimal'' theory \DelCR\ for \TCZ\ mentioned above. 
This theory is defined using a set of axioms (\BASIC{}
together with \OpenLIND), and the $\Delta_1^b$ {\em bit-comprehension rule}.
It is easy to see that \DelCR\ is a subset of both \rbarz and
$\mathbf{C}^0_2$, and
it follows from a result of Cook and Thapen \cite{Cook:Thapen:04}
that \DelCR\ is a proper subset of both unless the RSA encryption scheme
can be cracked in polynomial time.

The equational theories $A2V$ and $TV$ introduced by Johannsen
\cite{Johannsen:98:beam-buss} characterize \ACZ(2) and \TCZ, respectively.
These theories appear to be RSUV isomorphic respectively to our
second-order theories $\VZ(2)$ and \VTCZ.  One direction is clear:
the axioms of the equational theories translate to theorems of the
second-order theories.   To show the reverse direction
would require working out detailed proofs in the equational theories.

We show in Section~\ref{s:RSUV}
that \VTCZ\ is RSUV isomorphic to \DelCR.
From this and the previous paragraph
it appears that $TV$ and $\DelCR$ are equivalent.

\subsection{Second-Order Theories for \TCZ}
\label{s:sec-order-theories}

In \cite{Johannsen:Pollett:98}, the first-order theories 
$\mathbf{C}^0_{k+1}$
($k\ge 1$) are shown to be RSUV isomorphic to the second-order
theories $\mathbf{D}^0_k$. Thus, $\mathbf{D}^0_1$ can be seen as a
theory for \TCZ.  By the results of Cook and Thapen \cite{Cook:Thapen:04}
discussed above, $\mathbf{D}^0_1$ appears to be stronger than our theory
\VTCZ.

In \cite{Krajicek:95:feasibleII}, Kraj{\'{\i}}{\v{c}}ek introduces
the theory $(I\Sigma_0^{1,b})^{\mathit{count}}$ and notes that it should
correspond to constant-depth \fc, where \fc\ is an
extension of Frege proof systems.
It turns out that our theory \VTCZ\ is essentially the same as
$(I\Sigma_0^{1,b})^{\mathit{count}}$, but we note that  Kraj{\'{\i}}{\v{c}}ek
does not treat his theory in detail.

As argued in \cite{Cook:05:quaderni}, it seems that the second-order
logic used here is more appropriate for reasoning about small
complexity classes.  The usual first-order theories of bounded
arithmetic (including most of those described above for \TCZ)
include multiplication as a primitive operation, and include axioms
such as $x\cdot y = y\cdot x$.  Our second-order theories have no
primitive operations on second-order objects (strings) other than
length.  One advantage of this simplicity comes in the easy description
of the propositional translations of these theories \cite{Cook:05:quaderni}.
In order to show the RSUV isomorphism in Section \ref{s:RSUV}
between \VTCZ\ and \DelCR\ we must define binary multiplication
in \VTCZ\ and prove its properties, which is not an easy task.
But the alternative of simply assuming the commutative and distributive
laws as axioms is ``cheating'', rather like throwing in axioms for
the commutativity of multiplication in a propositional proof system for \TCZ.

\subsection{Organization}

Section \ref{s:SOL} presents the syntax and semantics of our second-order
theories, and defines the second-order versions of the complexity
classes \ACZ, \TCZ, \ACZ($m$), and \ACC.  Characterizations of
\TCZ\ and \ACZ($m$) are given in terms of threshold quantifiers and
modulo $m$ quantifiers, respectively.

Section \ref{s:theories} defines a finitely axiomatized theory
for each of the complexity classes mentioned in the introduction,
and introduces a universal conservative extension of each of these
theories which has function symbols for each function in the
associated class.  The main theorems state that the $\Sigma^B_1$-definable
functions in each theory are the functions in the associated
complexity class.  The theory \VTCZ\ for \TCZ\ is treated
in detail, and then a general method is introduced for defining theories
for other subclasses of \Ptime.  A general witnessing theorem is
proved.

Section \ref{s:RSUV} proves that our finitely-axiomatized
second-order theory \VTCZ\
is isomorphic to the first-order theory \DelCR\ of Johannsen
and Pollett.  It follows that there is a fixed upper bound of
the nesting depth of applications of the $\Delta^b_1$ bit-comprehension rule
required for proofs in \DelCR, which answers an open question
in \cite{Johannsen:Pollett:00}.

Section \ref{s:conclusion} summarizes our main contributions.
Appendix \ref{s:Interp-Addition} gives some details of the
RSUV isomorphism proof, and Appendix \ref{s:PHP} shows how the
proof of the Pigeonhole Principle can be formalized in \VTCZ.

\section{Second-Order Logic}
\label{s:SOL}

\subsection{Syntax and Semantics}
\label{s:Syntax-Semantics}

We use the two-sorted syntax of Zambella
\cite{Zambella:96:jsl, Zambella:97:apal} (see also
\cite{Cook:05:quaderni, Cook:02:notes}), which was inspired by Buss's
second-order theories defined in \cite{Buss:86:book}.
Our language has two
sorts of variables: the number variables $x, y, z, \ldots$
whose intended values are natural numbers; and
string variables $X, Y, Z, \ldots$, whose intended values are
finite sets of natural numbers (which represent binary strings).
Our two-sorted vocabulary $\LTwoA$ extends that
of Peano Arithmetic:
\[\LTwoA = [0,1,+,\cdot,|\ |; \in,\le,=^1, =^2].\]
Here $|\ |$ is a function from strings to numbers, and
the intended meaning of $|X|$ is 1 plus the largest element of $X$.
The binary predicate $\in$ denotes set membership. We will use
the abbreviation $X(t)$ for $t\in X$.
The equality predicates $=^1$ and $=^2$ are for numbers and strings,
respectively. We will write $=$ for
both $=^1$ and $=^2$; the exact meaning will be clear from the context.
The other symbols have their standard meanings.

{\em Number terms} are built from the constants 0,1, variables $x,y,z,...$,
and length terms $|X|$ using $+$ and $\cdot$.  We use $s,t,...$
for number terms.  The only {\em string terms}
are string variables $X,Y,Z,...$.   The atomic formulas are \True, \False,
(for True, False),
$s=t$, $X=Y$, $s\le t$, $t\in X$ for any number terms $s,t$ and string
variables $X,Y$.  Formulas are built from atomic formulas using
$\wedge,\vee, \neg$ and both number and string quantifiers
$\exists x, \exists X, \forall x,\forall X$.  Bounded number quantifiers
are defined as usual, and the bounded string quantifier
$\exists X \le t \  \varphi$ stands for $\exists X(|X|\leq t \wedge \varphi)$
and $\forall X\le t \ \varphi$ stands for $\forall X(|X|\le t\supset \varphi)$,
where $X$ does not occur in the term $t$.

A structure for $\LTwoA$  is defined in the same way as a structure
for a single-sorted language, except now there are two nonempty domains
$U_1$ and $U_2$, one for numbers and one for strings.  Each
symbol of $\LTwoA$ is interpreted in $\langle U_1,U_2\rangle$ by
a relation or function of appropriate type, with $=^1$ and $=^2$
interpreted as true equality on $U_1$ and $U_2$, respectively.
In the standard structure $\underline{\mathbb{N}}_2$, $U_1$ is $\N$ and
$U_2$ is the set of finite subsets of $\N$.  Each symbol of
$\LTwoA$ gets its intended interpretation.

In general we will consider a vocabulary $\mathcal{L}$ which extends
$\LTwoA$.
We require that the bounding terms (for the bounded quantifiers)
are restricted to mention the functions of \LTwoA\ only.
A formula is $\SigZB(\mathcal{L})$
if it has no string quantifiers and all number quantifiers are bounded.
A formula is $\SigOneB(\mathcal{L})$ ($\PiOneB(\mathcal{L})$,
$\SigOneOne(\mathcal{L})$, resp.)
if it is a $\SigZB(\mathcal{L})$ formula preceded by a block of
quantifiers of the form $\exists X\le t$ ($\forall X\le t$, $\exists X$, resp.).
If the block contains a single quantifier, the
formula is also called single-$\SigOneB(\mathcal{L})$
(single-$\PiOneB(\mathcal{L})$, single-$\SigOneOne(\mathcal{L})$, resp.).
A formula is $\gSigOneB(\mathcal{L})$ (resp. $\gPiOneB(\mathcal{L})$)
if it is obtained from $\SigZB(\mathcal{L})$ formulas using
the connectives $\wedge$ and $\vee$, bounded number
quantifiers and bounded existential (resp. universal) string
quantifiers (``{\bf g}'' for ``general'').
A formula is $\exists \gSigOneB(\mathcal{L})$ if it is of the form
$\exists \Xvec\varphi$, where $\varphi$ is $\gSigOneB(\mathcal{L})$.
We will omit $\mathcal{L}$ if it is \LTwoA.

The $\SigOneB$ formulas correspond to
(in first-order logic) strict $\SigOneb$ formulas
(i.e., $\SigOneb$ formulas where no bounded quantifier is inside the
scope of any sharply bounded quantifier),
while $\gSigOneB$ formulas correspond to $\SigOneb$ formulas.
Similar for $\PiOneB$ and $\gPiOneB$ formulas.

\subsection{Two-Sorted Complexity Classes}

We study two-sorted versions of standard complexity classes, where
the two sorts are those in the standard model $\underline{\mathbb{N}}_2$ for
$\LTwoA$:
the natural numbers and finite subsets of the natural numbers.
When a class is defined in terms of machines or circuits,
we assume that each number input is represented in unary notation
(i.e., $n$ is represented as a string of $n$ 1's),
and each finite subset $X$ is represented by its characteristic bit string.

There are two kinds of functions: {\em number functions} and {\em string functions}.
A number function $f(\xvec, \Xvec)$ takes values in \N, and a string function
$F(\xvec, \Xvec)$ take values in finite subsets of \N.
A function $f(\xvec, \Xvec)$ or $F(\xvec, \Xvec)$ is {\em polynomially bounded}
(or {\em p-bounded}) if there is a polynomial $\boldp(\xvec, \yvec)$ such that
$f(\xvec, \Xvec) < \boldp(\xvec, |\Xvec|)$, or $|F(\xvec, \Xvec)| < \boldp(\xvec, |\Xvec|)$.
The functions classes we consider here contain only p-bounded functions.

The class (uniform) \ACZ\ can be characterized as the set
of relations $R(\xvec, \Xvec)$ which
are accepted by
alternating Turing machines in time $O(\log(n))$ with constant alternations.
The following result is from \cite{Immerman:99:book, Cook:02:notes}.

\begin{theorem}
\label{t:AC0-rep}
A relation $R(\xvec,\Xvec)$ is in $\ACZ$ iff it is represented by
some $\SigZB$ formula $\varphi(\xvec,\Xvec)$.
\end{theorem}

We define \ACZ\ reducibility in the ``Turing'' style, as
opposed to the many-one style.  The idea is
(see for example \cite{Barrington:Immerman:Straubing:90}) that
$F$ is \ACZ\ reducible to $\calL$ if $F$ can be computed by a
(uniform) polynomial size constant depth family of circuits which
have unbounded fan-in gates computing functions from $\calL$, in
addition to Boolean gates.   We follow \cite{Cook:05:quaderni} and
make this precise in Definition \ref{d:AC0-red} below,
based on Theorem \ref{t:AC0-rep}.

The {\em bit graph} $B_F(z,\xvec,\Xvec)$ of a string function
$F(\xvec,\Xvec)$ is defined by the condition
\begin{equation}
\label{e:bitgraph}
B_F(z, \xvec, \Xvec) \equiv F(\xvec, \Xvec)(z)
\end{equation}

\begin{definition}
\label{d:sig0-def}
A string function is {\em $\SigZB$-definable} from a collection $\calL$
of two-sorted functions and relations if it is p-bounded and its bit graph is represented
by a $\SigZB(\calL)$ formula.
Similarly, a number function is {\em $\SigZB$-definable} from $\calL$ if it is p-bounded
and its graph is represented by a $\SigZB(\calL)$ formula.
\end{definition}

\begin{definition}
\label{d:AC0-red}
We say that a string function $F$ (resp. a number function $f$) is \ACZ\ reducible to $\calL$ if there is a
sequence of string functions $F_1, \ldots, F_n$ ($n \ge 0$) such that
\begin{equation}
\label{e:AC0-red}
F_i
\text{ is \SigZB-definable from }\calL \cup \{F_1, \ldots, F_{i-1}\},
\text{ for }i = 1, \ldots, n;
\end{equation}
and that $F$ (resp. $f$) is $\SigZB$-definable from $\calL \cup \{F_1, \ldots, F_n\}$.
A relation $R$ is $\ACZ$ reducible to $\calL$ if there is a
sequence $F_1, \ldots, F_n$ as above, and $R$ is represented by a $\SigZB(\calL \cup \{F_1, \ldots, F_n\})$
formula.
\end{definition}

The uniform classes \TCZ, $\ACZ(m)$, and \ACC\ can be defined in
several equivalent ways \cite{Barrington:Immerman:Straubing:90}.
Here we define them using \ACZ\
reducibility, and later we characterize them using generalized
quantifiers.
\begin{definition}
[{\TCZ, $\ACZ(m)$, \ACC}]
\label{d:TC0}
Let $\numones(z,X)$ be the number of elements of $X$ which are less than $z$.
\footnote{Thus the number of elements of $X$ is $\numones(|X|,X)$.}
Then \TCZ\ is the class of relations \ACZ\ reducible to $\numones$.
Similarly, for each $m \in \N$, $m \ge 2$,
let $\modulo_m(z,X) = (\numones(z,X) \mod m)$.
Then $\ACZ(m)$ is the class of relations \ACZ\ reducible to $\modulo_m$.
The class \ACC\ is the union of all $\ACZ(m)$, for $m \ge 2$.
\end{definition}

In general, each two-sorted relation class {\bf C} is associated with
a function class {\bf FC}.
\label{d:FC}
A number function belongs to {\bf FC} if it is p-bounded, and its graph is in {
\bf C}.
A string function $F(\xvec, \Xvec)$ belongs to {\bf FC} if it is p-bounded,
and its bit graph (\ref{e:bitgraph}) is in {\bf C}.

\begin{lemma}
\FTCZ\ is the class of functions \ACZ\ reducible to \numones.
For $m \in \N, m \ge 2$, $\FACZ(m)$ is the class of functions \ACZ\ reducible to $\modulo_m$.
\FACC\ is the class of functions \ACZ\ reducible to $\modulo_m$, for some $m$.
\end{lemma}

Another characterization of \TCZ\ is as follows.
Consider augmenting the current two-sorted logic with the
{\em counting} quantifier, i.e.,
\begin{equation*}
\exists s\ z < t \varphi(z, \xvec, \Xvec)
\end{equation*}
is a formula which is true if and only if there are exactly $s$ values of $z$
such that $\varphi(z, \xvec, \Xvec)$ is true.
It has been shown \cite[Theorem 2.9]{Nguyen:04:thesis} that \TCZ\ is exactly the class of relations
represented by \SigZBCount\ formulas, i.e., bounded formulas which allow only number quantifiers
and the counting quantifier.
The proof in \cite{Nguyen:04:thesis} is based on the characterization of
\TCZ\ \cite{Barrington:Immerman:Straubing:90}: $\TCZ = \FOCount$.
It shows how to translate \FOCount\ formulas into
\SigZBCount\ formulas, and vice versa.

\subsection{The Threshold Quantifier and Threshold Operation}

Observe that the counting quantifier as discussed above ``counts''
{\em exactly} 
the number of $z$'s that make $\varphi(z)$ true.
We now define the
{\em threshold} quantifier, which has syntax
\begin{equation}
\label{e:thresh-quant}
       \thq s\ z<t \varphi(z)
\end{equation}
where $s,t$ are terms not containing $z$.  (The variable $z$ is
bound by the quantifier.)
The semantics is given by the condition that (\ref{e:thresh-quant})
holds  if and only if there are {\em at least} $s$
values of $z$ less than $t$ that make $\varphi$ true.
This is similar to the counting quantifier, but
we find the threshold quantifier more convenient, and will use it here.

Let \SigZBTh\ be the class of formulas built in the same way as
$\SigZB$, except now we allow threshold quantifiers in addition
to bounded number quantifiers. 
The following result and its corollary provide interesting
characterizations of \TCZ, but they are not used in the rest
of this paper.

\begin{theorem}
\label{t:TC0-threshold-quant}
\TCZ\ is the class of relations represented by \SigZBTh\ formulas.
\end{theorem}

\begin{proof}
First, let $\varphi(\xvec, \Xvec)$ be a \SigZBTh\ formula.
We will prove by induction on the structure of $\varphi$ that it represents
a \TCZ\ relation.
The base case where $\varphi$ is an atomic formula is straightforward.
For the induction step, consider the interesting case where
\[\varphi(\xvec, \Xvec) \equiv \thq s\ z < t\ \varphi'(z, \xvec, \Xvec).\]
By the induction hypothesis, $\varphi'(z, \xvec, \Xvec)$ represents a \TCZ\ relation.
In other words, it represents the same relation as some $\SigZB(\{\numones, F_1, \ldots, F_n\})$
formula $\psi(z, \xvec, \Xvec)$, for a sequence $F_1, \ldots, F_n$
of string functions
satisfying \eqref{e:AC0-red}.
Let $F_{n+1}$ be \SigZB-definable from $\{\numones, F_1, \ldots, F_n\}$ as follows:
\[F_{n+1}(\xvec, \Xvec) (z) \Lra z < t \wedge \psi(z, \xvec, \Xvec).\]
Then $\varphi$ represents the same relation as the formula
\[ \numones(t,F_{n+1}(\xvec, \Xvec)) \ge s.\]

For the other direction, we will prove by induction on $n \ge 0$
a stronger result:

{\em If $F_1, \ldots, F_n$ is any sequence of string functions
satisfying \eqref{e:AC0-red} (with ${\calL}=\{\numones\}$),
then for any
$\SigZBTh(\{\numones, F_1, \ldots, F_n\})$ formula
$\psi(\xvec, \Xvec)$
there is a $\SigZBTh$ formula $\varphi(\xvec, \Xvec)$ that
represents the same relation.}
\hfill(*)

a) For the base case, we prove by induction on the structure of a
$\SigZBTh(\numones)$ formula
$\psi$ that there is a $\SigZBTh$ formula $\varphi$ that represents the
same relation as $\psi$.
It suffices to show that any atomic formula $\psi(\numones)$ (i.e., $\psi$ contains \numones)
is equivalent to a $\SigZBTh$ formula.
Let
$$\numones(t_1,X_1), \ldots, \numones(t_m,X_m)$$
be all
occurrences of $\numones$ in $\psi$, enumerated in some order such that
if $\numones(t_i,X_i)$ is a sub-term of $t_j$ then $i<j$.
Let $u_1,...,u_m$ be a list of new variables.  Let $t'_j$ be
the result of replacing each maximal sub-term $\numones(t_i,X_i)$
of $t_j$ by $u_i$ and let $\psi'$ be the result of replacing each
maximal sub-term $\numones(t_i,X_i)$ in $\psi$ by $u_i$.
Note that if $u_i$ occurs in $t'_j$ then $i<j$.
Now the $\SigZBTh$ formula $\varphi$ is
\begin{equation*}
\exists u_1 \le t'_1 \ldots \exists u_m \le t'_m,\ \psi' \wedge
\bigwedge_{k = 1}^m [\thq u_k\ z < t'_k\ X_k(z)
\wedge \neg \thq (u_k+1)\ z < t'_k\ X_k(z)].
\end{equation*}

b) For the induction step, suppose that $F_{n+1}$ is $\SigZB$-definable from
$\{\numones, F_1, ..., F_n\}$ (for $n \ge 0$).
Let $\psi$ be a $\SigZBTh(\{\numones, F_1, \ldots, F_{n+1}\})$ formula.
We will show how to eliminate $F_{n+1}$ from $\psi$ by induction
on the depth of nesting of $F_{n+1}$ in $\psi$.

For a term or formula $\omega$, we define $d(\omega)$ to be the
maximum depth of nesting of any occurrence of $F_{n+1}$ in $\omega$.

We will prove the following by induction on $k \ge 0$:

{\em If $d(\psi) = k$, then there is a $\SigZBTh$ formula $\varphi$ that
represents the same relation as $\psi$.}
\hfill(**)

(i) The base case where $k = 0$ follows from the induction hypothesis of (*),
since there is no occurrence of $F_{n+1}$ in $\psi$.

(ii) Suppose that (**) holds for all $\psi$ where $d(\psi) \le k$.
It suffices to prove (**) when $\psi$ is an atomic formula, and $d(\psi) = k + 1$.

Let $F_{n+1}(\rvec_1, \Tvec_1), \ldots, F_{n+1}(\rvec_\ell, \Tvec_\ell)$
be all string terms in $\psi$ of the form $F_{n+1}(\rvec, \Tvec)$,
where $d(F_{n+1}(\rvec, \Tvec)) = k + 1$.
(Thus $\max(\{d(r)\mid r \in \rvec_i\} \cup \{d(T) \mid T \in \Tvec_i\}) = k$,
for $i = 1, \ldots, \ell$.)
Let $W_1,...,W_\ell$ be new string variables, and
let $\psi'$ be the formula obtained from $\psi$ by replacing each
$F_{n+1}(\rvec_i, \Tvec_i)$ with $W_i$, $i=1,...,\ell$.  
Then $\psi'$ is atomic, since $\psi$ is atomic.
Since $d(\psi') \le k$, it follows
by the induction hypothesis that there is a $\SigZBTh$ formula $\theta$
that represents the same relation as $\psi'$.
Then each $W_i$ can occur in $\theta$ only in the form $|W_i|$,
or $W_i(r)$, for some number term $r$ ($r$ might contain some $|W_j|$'s).

Suppose that $F_{n+1}$ is defined by 
$$
   F_{n+1}(\xvec,\Xvec)(z) \  \lra \  z<t_{n+1}(\xvec,\Xvec) \wedge
                \varphi_{n+1}(z,\xvec,\Xvec)
$$
where $t_{n+1}$ is a term in the base language $\LTwoA$ and
$\varphi_{n+1}$ is a $\SigZB(\{\numones,F_1,...,F_n\})$ formula.
Now each occurrence of $|W_i|$ in $\theta$ can be eliminated by
the equivalence
$$
   |F_{n+1}(\rvec_i, \Tvec_i)| = z_i \  \lra \  \delta_i
$$
where $z_1,...,z_\ell$ are new number variables and
(setting $t'_i\equiv t_{n+1}(\rvec_i,\Tvec_i)$)
\[
    \delta_i \  \equiv \  z_i \le t'_i \wedge 
[\forall x < t'_i, \ x \ge z_i \supset \neg \varphi_{n+1}(x, \rvec_i, \Tvec_i)]
\wedge
[z_i > 0 \supset \varphi_{n+1}(z_i-1, \rvec_i, \Tvec_i)]
\]
(Note that $d(\varphi_{n+1}(z, \rvec_i, \Tvec_i)) \le k$.)
Let $\theta'$ be obtained from $\theta$ by replacing each $|W_i|$ by $z_i$.
Let $\theta''$ be the formula
\[
\exists z_1 \le t'_1 \ldots \exists z_\ell \le t'_\ell 
     (\theta'\wedge\delta_1\wedge...\wedge\delta_\ell)
\]
Next, replace each occurrence of the form $W_i(r)$ in $\theta''$
(such $r$ does not contain any of the $W_j$'s) with
\[r < t'_i \wedge \varphi_{n+1}(r, \rvec_i, \Tvec_i).\]
Let $\theta'''$ be the resulting formula.
Then $\theta'''$ represents the same relation as $\psi$.
Since $d(\theta''') \le k$, we can apply the induction hypothesis
to $\theta'''$ to obtained the desired $\SigZBTh$ formula $\varphi$
that represents the same relation as $\psi$.
\end{proof}

Define the {\em threshold} operation as follows.
It takes a relation $Q(z)$ (which may contain other parameters)
to the relation $[\Th zQ](k, y)$ defined by
\[[\Th zQ](k, y) \Lra \mbox{ there are at least } k \mbox{ values of } z < y \mbox{ that satisfy } Q(z).\]
Then the threshold and Boolean operations together simulate
the bounded number quantification operations.
For example, the relation $[\exists z \le t Q](z)$ is the same as the relation
\[[\Th z Q](1, t).\]

The following is immediate from Theorem~\ref{t:TC0-threshold-quant}.
\begin{corollary}
\label{t:TC0-threshold-oper}
\TCZ\ is the closure of \ACZ\ relations under the threshold and Boolean operations.
\end{corollary}

\subsection{The Modulo $m$ Quantifier and Operation}

For each $m \in \N, m \ge 2$, the {\em modulo $m$ quantifier} and 
{\em modulo $m$ operation} can be defined similarly as the
threshold quantifier and threshold operation, with a little more complication
(see \cite{Paris:Wilkie:85}).
In particular, the modulo $m$ quantifier $\Mod_m$ only makes sense when the
variable it quantifies over is bounded.
Thus, 
\[\Mod_m z < t\ \varphi(z)\]
 is true if and only if the number of
$z < t$ satisfying $\varphi(z)$ is exactly 1 modulo $m$.
Similarly, the modulo $m$ operation takes a relation $Q(z, \xvec, \Xvec)$
into the relation $[\Mod_m z Q](y, \xvec, \Xvec)$ which consists of
all tuples $(y, \xvec, \Xvec)$ such that
\[|\{(z, \xvec, \Xvec): z < y \mbox{ and } Q(z, \xvec, \Xvec)\}| = 1 \mod{m}.\]

Let \SigZBModm\ formulas be bounded formulas in our two-sorted logic
augmented with
the $\Mod_m$ quantifier, where only bounded number quantifiers are allowed.
Then the analog of Theorem~\ref{t:TC0-threshold-quant} and
Corollary~\ref{t:TC0-threshold-oper} can be proved by slight modifications
of the original proofs.

\begin{theorem}
For each $m \ \in \N, m \ge 2$, $\ACZ(m)$ is the class of relations represented
by \SigZBModm\ formulas.
It is also the closure of \ACZ\ relations under Boolean, bounded number quantification
and modulo $m$ operations.
\end{theorem}

\section{The Theories}
\label{s:theories}

\subsection{The Theory \VZ}

We start by describing the theory \VZ\ \cite{Cook:02:notes, Cook:05:quaderni}
for the complexity class \ACZ.
All of the theories that we introduce here are extensions of \VZ.

The theory \VZ\ has underlying language $\LTwoA$ and is axiomatized by
the set of axioms \BASIC{2} and the \COMP{\SigZB} axiom scheme.
First, \BASIC{2}%
\label{d:BASIC}
is the set of the axioms {\bf B1 -- B12}, {\bf L1, L2} and {\bf SE} below.

\begin{figure}[htb]
\centering
\begin{tabular}{|ll|}
\hline
{\bf B1.} $x+1\neq 0$                             & {\bf B7.} $(x\le y \wedge y\le x)\supset x = y$\\
{\bf B2.} $x+1 = y+1\supset x=y$                  & {\bf B8.} $0\le x$\\
{\bf B3.} $x+0 = x$                               & {\bf B9.} $x\le y \wedge y\le z\supset x\le z$\\
{\bf B4.} $x+(y+1) = (x+y) + 1$                   & {\bf B10.} $x\le y \vee y\le x$\\
{\bf B5.} $x\cdot 0 = 0$                          & {\bf B11.} $x\le y \lra x< y+1$\\
{\bf B6.} $x\cdot (y+1) = (x\cdot y) + x$         & {\bf B12.} $x\neq 0 \supset \exists y<x(y+1 = x)$\\
{\bf L1.} $X(y) \supset y < |X|$                  & {\bf L2.}  $y+1 = |X| \supset X(y)$ \\
\multicolumn{2}{|l|}
{{\bf SE.} $X = Y \lra\ [|X| = |Y|\wedge\forall z<|X|(X(z)\lra Y(z))]$}  \\
\hline
\end{tabular}
\end{figure}

The axiom scheme \COMP{\SigZB} is the set of all formula of the form
\begin{equation}
\label{e:SBcomp}
\exists X \le a \forall z < a,\ X(z) \lra \varphi(z),
\end{equation}
where $\varphi$ is a $\SigZB$ formula not containing $X$.

Although $\VZ$ does not have an explicit induction scheme, axioms
{\bf L1} and {\bf L2} tell us that if $X$ is nonempty then it has a largest
element, and thus we can show that $\VZ$
proves the $\MIN{X}$ formula
$$   0<|X|\supset\exists x<|X|(X(x)\wedge \forall y<x \ \neg X(y))  $$
and $\IND{X}$
\[
  [X(0) \wedge \forall y < z(X(y) \supset X(y+1))] \supset X(z)
\]
From this and \COMP{\SigZB} we conclude that
$\VZ$ proves the scheme
$$  \IND{\SigZB}\mbox{:} \qquad
[\varphi(0)\wedge\forall x(\varphi(x)\supset \varphi(x+1))]
      \supset \forall z\varphi(z)
$$
where $\varphi(x)$ is any $\SigZB$ formula (possibly containing parameters).

A pairing function can be defined in \VZ\ by using
 $\langle x,y\rangle$ to abbreviate the term
$(x+y)(x+y+1)+2y$.  Then \VZ\
proves that the map $(x,y) \mapsto \langle x,y\rangle$
is an one-one map from $\N\times\N$ to $\N$.
We use this idea to define a binary array
$X$ using the definition $X(x,y) \equiv X(\langle x,y\rangle)$.
By iterating the pairing function we can define a multidimensional
array $X(\xvec)$.  Then \VZ\ proves the corresponding comprehension
scheme
$$
   \exists Z\le \langle \avec\rangle \forall \zvec < \langle \avec\rangle,
                Z(\zvec)\lra \varphi(\zvec)
$$
for any $\SigZB$ formula $\varphi$.

If we think of $Z$ as a two-dimensional array, then we can
represent row $x$ in this array by $Z^{[x]}$, 
where $ Z^{[x]}=\Row(x,Z)$ is the $\FACZ$ string function with
bit-defining axiom
\begin{equation}
\label{e:Row-def}
    Z^{[x]}(i) \lra \Row(x,Z)(i) \lra i<|Z|\wedge Z(x,i)
\end{equation}

\begin{lemma}
\label{l:row-pairing}
Let $\VZ(\Row)$ be the extension of \VZ\ obtained by adding the function
$\Row$ with defining axiom (\ref{e:Row-def}).  Then $\VZ(\Row)$
is conservative over \VZ, and every $\SigZB(\Row)$ formula $\varphi$
is provably equivalent in $\VZ(\Row)$ to a \SigZB\ formula $\varphi'$.
\end{lemma}

\begin{proof}
Conservativity follows from the fact that $\Row$ is \SigOneB-definable
in \VZ (see Lemma \ref{l:sigma-cons}). 

For the second part, we may assume by the axiom {\bf SE} that
$\varphi$ does not contain $=^2$.  We proceed by induction on the
maximum nesting depth of $\Row$ in $\varphi$.  It suffices to consider
the case in which $\varphi$ is atomic.  If $\varphi$ has the
form $\Row(t,T)(s)$, then $\varphi$ is equivalent to $T(t,s)$ which
by the induction hypothesis is equivalent to a \SigZB\ formula.

Now suppose that $\varphi$ is atomic and does not have the form $\Row(t,T)(s)$.
Let
$$\Row(t_1,T_1),...,\Row(t_k,T_k)$$
be the maximal depth string terms
occurring in $\varphi$.  Then each such term $\Row(t_i,T_i)$ must occur
in the context $|\Row(t_i,T_i)|$, so
$$
   \varphi\equiv \varphi'(|\Row(t_1,T_1)|,...,|\Row(t_k,T_k)|)
$$
where $\varphi'(x_1,...,x_k)$ has less $\Row$-nesting depth than $\varphi$.
Then $\VZ(\Row)$ proves
$$
   \varphi \lra \exists x_1\le |T_1|...\exists x_k\le |T_k|,
      (\bigwedge_{i=1}^k row\mbox{-}length(x_i,t_i,T_i)) \wedge \varphi(x_1,...,x_k)
$$
where $row$-$length(x,y,Z)$ is a \SigZB\ formula expressing the condition
$x=|\Row(y,Z)|$.  We can now apply the induction hypothesis to the RHS.
\end{proof}


\subsection{The Theory \VTCZ}

The theory \VTCZ\ is \VZ\ together with \NUMONES, which is essentially
a \SigOneB\ defining axiom for \numones\ (Definition \ref{d:TC0}) .
Let $\varphi_\NUMONES(X, Y)$ be the \SigZB\ formula stating that $Y$
is a counting array of $X$, i.e., for each $z \le |X|$, $Y(z, y)$ holds
if and only if $\numones(z,X) = y$:
\begin{multline}
\label{e:NUMONES}
\varphi_\NUMONES(X,Y)\equiv 
 \ [\forall z\le |X| \exists! y\le |X| Y(z,y)]\ \wedge \
Y(0,0)\ \wedge\\ 
 \forall z<|X| \forall y\le |X|,\
Y(z,y) \supset [(X(z)\supset Y(z+1,y+1)) \wedge
(\neg X(z)\supset Y(z+1,y))].
\end{multline}

\begin{definition}
Let \NUMONES\ denote
$\forall X \exists Y \varphi_\NUMONES(X,Y)$.
The theory $\VTCZ$ is $\VZ$ extended by the axiom \NUMONES.
\end{definition}

Note that \VZ\ proves that \NUMONES\ implies that same axiom with $\exists Y$
replaced by the bounded quantifier $\exists Y\le 1 + \langle |X|, |X| \rangle$.
Hence \VTCZ\ is equivalent to a theory with bounded axioms.

\ignore{
Theories for the complexity classes $\ACZ(m)$ and $\ACC$ are defined
in the same way that \VTCZ\ is defined for the class \TCZ.
For $m \ge 2$, $\varphi_{\MOD_m}(X, Y)$ is the formula stating that $Y$ is
the ``counting modulo $m$'' array for $X$:
\begin{equation}
\label{e:MOD-m}
\begin{split}
\varphi_{\MOD_m}(X,Y)\equiv &
[\forall z\le |X| \exists! y < m Y(z,y)]\ \wedge\ 
Y(0,0)\ \wedge\
\forall z<|X| \forall y < m,\\
& Y(z,y)\supset [(X(z)\supset Y(z+1, \ y+1 \mod{m})) \wedge
(\neg X(z)\supset Y(z+1,y))].
\end{split}
\end{equation}
Here, we identify the natural number $m$ with the corresponding numeral $\numm$.
We take $\varphi(y \mod{m})$ as an abbreviation for 
\begin{equation}
\label{e:mod-m}
\exists r < m, \exists q \le y,\ y = qm + r \wedge \varphi(r).
\end{equation}
Thus if $\varphi(y)$ is \SigZB, then $\varphi(y\mod{m})$ is also \SigZB.

\begin{definition}
For each $m\ge 2$, let 
$\MOD_m \equiv \forall X \exists Y \varphi_{\MOD_m}(X,Y)$.  Then
\begin{eqnarray*}
\VZ(m) & = & \VZ \cup \{\MOD_m\}  \\
\VACC & = & \VZ \cup \{\MOD_m \mid m \ge 2 \}.
\end{eqnarray*}
\end{definition}
Note that the string $Y$ in $\MOD_m$ can be bounded by
$\langle |X|, m \rangle$.
}

\begin{lemma}
The theories \VZ\ and \VTCZ\ are finitely axiomatizable.
\end{lemma}

\begin{proof}
The finite axiomatizability of \VZ\ is proved in \cite{Cook:Kolokolova:03}.
The theory \VTCZ\ is the result of adding a single axiom to \VZ.
\end{proof}

The next definition refers to the notion of $\SigOneOne(\mathcal{L})$
formula, defined in Section \ref{s:Syntax-Semantics}.

\begin{definition}\label{d:Phi-definable}
Let $\mathcal{T}$ be an extension of $\VZ$ over a language $\mathcal{L}$.
A string function $F(\xvec,\Xvec)$
is $\Sigma_1^1(\mathcal{L})$-{\em definable} in $\mathcal{T}$ if it satisfies
\begin{equation}\label{e:function-intro}
    Y=F(\xvec,\Xvec)\lra \varphi(\xvec,\Xvec,Y)
\end{equation}
for some $\Sigma_1^1(\mathcal{L})$ formula $\varphi$, and
\begin{equation}\label{e:uniqueness}
   \mathcal{T}\vdash \forall\xvec\forall\Xvec\exists ! Y\phi(\xvec,\Xvec,Y)
\end{equation}
The $\Sigma_1^1(\mathcal{L})$-definability for a number function
$f(\xvec,\Xvec)$ is defined similarly.
\end{definition}

\begin{lemma}
\label{l:sigma-cons}
If $\mathcal{T}$ is an extension of \VZ\ which satisfies
(\ref{e:uniqueness}) and $F$ is not in the language of $\mathcal{T}$
and $\mathcal{T}'$ is the result of adding $F$ to the language and
adding (\ref{e:function-intro}) as an axiom, then $\calT'$ is
a conservative extension of $\mathcal{T}$.
\end{lemma}

\begin{proof}
According to (\ref{e:uniqueness}), every model of $\mathcal{T}$ has
an expansion to a model of $\mathcal{T}'$ which satisfies
(\ref{e:function-intro}).
\end{proof}

If $\mathcal{T}$ is a bounded theory, in the sense that the quantifiers
in the axioms for $\mathcal{T}$ can be bounded by terms of $\LTwoA$,
then by Parikh's Theorem \cite{Parikh:71, Cook:02:notes} it follows that a
function is $\Sigma_1^1(\mathcal{L})$ definable in $\mathcal{T}$ iff it is
$\Sigma^B_1$ definable in $\mathcal{T}$.

We can now state one of our main results, which explains the sense in
which our theories characterize the corresponding complexity classes.
We already know \cite{Cook:05:quaderni} that the $\Sigma^1_1$-definable
(and hence the $\Sigma_1^B$)-definable functions in \VZ\
are precisely those in \FACZ.

\begin{theorem}
\label{t:definable-fcns}
The $\Sigma_1^1$-definable (and the $\Sigma_1^B$-definable)
functions in \VTCZ 
are precisely those in \FTCZ. 
\end{theorem}

The proof is the subject of Subsections
\ref{s:universal} - \ref{s:wit-thms}.

\ignore{
\begin{corollary}
If \VACC\ is finitely axiomatizable, then $\ACC = \ACZ(m)$, for some $m$.
\end{corollary}

\begin{proof}
If \VACC\ is finitely axiomatizable, then by compactness, it is equal to
\[\VZ \cup \{\MOD_i \mid 2 \le i \le m' \},\]
for some $m'$.
Let $m = \lcm\{2, \ldots, m'\}$, then
\[\VZ(m) \vdash \{\MOD_i \mid 2 \le i \le m' \}.\]
Therefore $\VACC = \VZ(m)$, and the conclusion follows from the theorem.
\end{proof}

\begin{corollary}
If $\VACC \vdash \NUMONES$, then $\TCZ = \ACZ(m)$, for some $m$.
\end{corollary}
}

\subsection{Universal Theories}
\label{s:universal}

We will employ the techniques from \cite{Cook:05:quaderni}
to develop the universal version of our theories.
The idea is to introduce Skolem functions which are provably total
in the theories to eliminate the quantifiers.
Note that the axioms {\bf B12} and {\bf SE} are not universal statements.
As in \cite{Cook:05:quaderni}, {\bf B12} is replaced by 
$\mathbf{B12}'$ and $\mathbf{B12}''$ below.
Consider the number function \pd\ where $\pd(x)$ is the predecessor of $x$.
Then $\mathbf{B12}'$ and $\mathbf{B12}''$ are the defining axioms of \pd:
\begin{align}
\label{e:pd}
{\mathbf{B12}'}\ \pd(0) = 0         \qquad
{\mathbf{B12}''}\ x\neq 0 \supset \pd(x) + 1 = x
\end{align}

The left-to-right direction of {\bf SE} can be expressed by an
open formula simply by replacing $\forall z < |X|$ by $z < |X| \supset$:
\begin{equation*}
\notag
\mathbf{SE}': X = Y \supset [|X| = |Y| \wedge z < |X| \supset (X(z) \lra Y(z))].
\end{equation*}
The right-to-left direction of {\bf SE} has an implicit quantifier $\exists z < |X|$.
We can get rid of this by using the function $\fSE$
\label{d:fSE}%
(which is $\falphat$ in Definition~\ref{d:LFACZ} below, when $\alpha \equiv X(z) \not\lra Y(z)$, and $t = |X|$):
\begin{align}
\label{e:fSE-1}
& \fSE(X, Y) \le |X|,\\
\label{e:fSE-2}
& z < |X| \wedge (X(z) \not\lra Y(z)) \supset X(\fSE(X, Y)) \not\lra Y(\fSE(X, Y)), \\
\label{e:fSE-3}
& z < \fSE(X, Y) \supset X(z) \lra Y(z).
\end{align} 
Thus $\fSE(X, Y)$ is the smallest number $< |X|$ which distinguishes $X$ and $Y$, and
$|X|$ if no such number exists.
Let {\bf SE}$''$ be
\begin{equation*}
\notag
\mathbf{SE}'': (|X| = |Y| \wedge \fSE(X, Y) = |X|) \supset X = Y.
\end{equation*}

\begin{definition}
[{\LFACZ}]
\label{d:LFACZ}
\LFACZ\ is the smallest class that satisfies\\
{\em a)}
\LFACZ\ includes $\LTwoA \cup \{\pd, \fSE\}$.\\
{\em b)}
For each open formula $\alpha(z,\xvec,\Xvec)$ over $\LFACZ$ and term
$t=t(\xvec,\Xvec)$ of $\LTwoA$ there is a string function $F_{\alpha,t}$
of $\LFACZ$ with defining axiom
\begin{equation}
\label{e:LFACZ-F}
  F_{\alpha,t}(\xvec,\Xvec)(z) \lra z<t\wedge\alpha(z,\xvec,\Xvec)
\end{equation}
{\em c)}
For each open formula $\alpha(z,\xvec,\Xvec)$ over $\LFACZ$ and term
$t=t(\xvec,\Xvec)$ of $\LTwoA$ there is a number function $f_{\alpha,t}$
with defining axioms
\begin{align}
   &  f_{\alpha,t}(\xvec,\Xvec) \le t(\xvec,\Xvec) \label{e:LFACZ-f-1}  \\
   &   f_{\alpha,t}(\xvec,\Xvec)<t(\xvec,\Xvec)\supset
                   \alpha(f_{\alpha,t}(\xvec,\Xvec),\xvec,\Xvec))
            \label{e:LFACZ-f-2} \\
   &  z<f_{\alpha,t}(\xvec,\Xvec)\supset \neg\alpha(z,\xvec,\Xvec)
                                                           \label{e:LFACZ-f-3}
\end{align}
\end{definition}

Note that
$f_{\alpha,t}(\xvec,\Xvec) = \min z<t \, \alpha(z,\xvec,\Xvec)$
and
\begin{equation}
\label{e:quant-define}
   \exists z<t \, \alpha(z,\xvec,\Xvec) \lra
         f_{\alpha,t}(\xvec,\Xvec) < t.
\end{equation}


We define the theory \VZbar\ \cite{Cook:05:quaderni} to be the universal
theory over \LFACZ\ whose
axioms are the universal closures of the following list of open
formulas: {\bf B1 - B11}, {\bf B12$'$}, {\bf B12$''$},
{\bf L1}, {\bf L2}, {\bf SE$'$}, {\bf SE$''$}, the defining axioms
\eqref{e:fSE-1}, \eqref{e:fSE-2}, \eqref{e:fSE-3} for $\fSE$,
and the defining axiom \eqref{e:LFACZ-F} for each function $\Falphat$
and defining axioms \eqref{e:LFACZ-f-1}, \eqref{e:LFACZ-f-2}, \eqref{e:LFACZ-f-3}
for each function \falphat.

\begin{lemma}
\label{t:open=sigma}
For every $\SigZB$ formula $\varphi$ there is an open formula $\alpha$
of \LFACZ\ such that $\VZbar$ proves $(\varphi\lra\alpha)$.
For every open formula $\alpha$ of $\LFACZ$ there is a $\SigZB$ formula
$\varphi$ such that $\VZbar$ proves $(\varphi\lra\alpha)$.
\end{lemma}

\begin{proof}
The first sentence follows by structural induction on $\SigZB$ formulas
$\varphi$, using (\ref{e:quant-define}).  To prove the second sentence
consider an enumeration of the new function symbols of $\LFACZ$
in some order such that the defining axioms of each function in
the list mention only earlier functions in the list.  Now show
by induction on $k$ that if $\alpha$ only involves functions
occurring in the first $k$ positions on the list then $\alpha$
is equivalent to some $\SigZB$ formula $\varphi$. 
\end{proof}

\begin{theorem}\label{t:Vconserve}
$\VZbar$ is a conservative extension of $\VZ$.  The function
symbols in $\LFACZ$ represent precisely the functions in $\FACZ$.
\end{theorem}

\begin{proof}
To show that $\VZbar$ extends \VZ\ it suffices to show
$\VZbar$ proves the \COMP{\SigZB} axioms (\ref{e:SBcomp}).
From the first sentence of Lemma \ref{t:open=sigma} and
(\ref{e:LFACZ-F}) we have that for
every $\SigZB$ formula $\varphi(z,\xvec,\Xvec)$
there is a function $F$ in $\LFACZ$ such that
$$
   \VZbar \vdash \ \ F(\xvec,\Xvec)(z) \lra z<a\wedge \varphi(z,\xvec,\Xvec)
$$
from which (\ref{e:SBcomp}) follows.

To see that the extension is conservative we can prove by induction
that the functions in \LFACZ\ are \SigOneB-definable in \VZ.
Here we enumerate \LFACZ\ so that each function is defined
from earlier functions in the enumeration, starting with \pd, \fSE\ and \Row.
The main step is to show that the quantifier-free defining axiom for the
$(n+1)$-st function can be translated into a $\SigOneB(\LTwoA)$
defining axiom in \VZ.
Finally it is clear from
Definition \ref{d:LFACZ} that the function symbols
in $\LFACZ$ represent precisely the functions in $\FACZ$.
\end{proof}

It is worth emphasizing that $\VZbar$ proves the \IND{\SigZB}
and \MIN{\SigZB} schemes, since it extends $\VZ$.  This is true even
though $\VZbar$ has purely universal axioms, and has no explicit
induction axiom or rule.

Below we prove the General Witnessing Theorem for universal theories
(Theorem~\ref{t:witnessing}).
The Witnessing Theorems for our theories will follow from those of
their corresponding universal conservative extensions.
It follows that the $\exists \gSigOneB$-definable functions in these theories
are in the appropriate complexity classes.  For the other direction,
it is clear that the universal theories define all functions in the
appropriate classes, and Theorem \ref{t:sig0-comp-new-syms}
below shows the same for the original theories.

\subsection{The Theory \VTCZbar}
\label{s:vtcz}

The function \numones\ from Definition \ref{d:TC0} has defining
axioms
\label{d:numones}
\begin{align}
\label{e:numones-1}
\numones(0,X) = 0 \\
\label{e:numones-2}
X(z)  \supset  \numones(z+1,X) = \numones(z,X) + 1 \\
\label{e:numones-3}
\neg X(z)  \supset  \numones(z+1,X) = \numones(z,X).
\end{align}

Since 
\begin{equation}
\label{e:numones-NUM}
   y=\numones(x,X) \lra \exists Y, \ \varphi_\NUMONES(X,Y) \wedge Y(x,y)
\end{equation}
it is easy to see that \numones\ is \SigOneB-definable in \VTCZ.

The vocabulary \LFTCZ\ includes \numones\ and
is intended to represent the functions in \FTCZ.

\begin{definition}
\label{d:LFTCZ}
\LFTCZ\ is defined in the same way as \LFACZ\ (Definition~\ref{d:LFACZ})
with (a), (b) and (c) replaced by\\
(a$'$) \LFTCZ\ includes $\LTwoA \cup \{\pd, \fSE, \numones\}$,\\
(b $'$), (c$'$) are the same as (b), (c), except that
\LFACZ\ is replaced by \LFTCZ.
\end{definition}

The next lemma follows directly from the definitions.

\begin{lemma}
\label{t:LFTCZ-repre}
The functions in \LFTCZ\ represent precisely \FTCZ.
A relation is in \TCZ\ if and only if it is represented by some open $\LFTCZ$ formula.
\end{lemma}

\begin{definition}
\label{d:VTCZbar}
\VTCZbar\ is the universal theory over \LFTCZ\ whose axioms
are the universal closures of the following list of open formulas:
{\bf B1 - B11}, {\bf B12$'$}, {\bf B12$''$},
{\bf L1}, {\bf L2}, {\bf SE$'$}, {\bf SE$''$},
the defining axioms \eqref{e:fSE-1}, \eqref{e:fSE-2}, \eqref{e:fSE-3} for $\fSE$,
the defining axioms \eqref{e:numones-1}, \eqref{e:numones-2}, \eqref{e:numones-3} for $\numones$,
and the defining axioms
\eqref{e:LFACZ-F} and \eqref{e:LFACZ-f-1}, \eqref{e:LFACZ-f-2}, \eqref{e:LFACZ-f-3}
for the functions of \LFTCZ.
\end{definition}

The first part of the analog of Lemma \ref{t:open=sigma} is easily shown
to hold in this context.

\begin{lemma}
\label{t:open-numones}
For every $\SigZB(\numones)$ formula $\varphi$ there is an open formula $\alpha$
of \LFTCZ\ such that \VTCZbar\ proves $(\varphi\lra\alpha)$.
\end{lemma}

From this we can show the following.

\begin{lemma}
\label{t:VTCZ-extend}
\VTCZbar\ extends \VTCZ.
\end{lemma}

\begin{proof}
Since \VTCZbar\ extends \VZbar, it suffices to show \VTCZbar\
proves \NUMONES.  We show this by pointing out that \LFTCZ\
includes a string function $F_{NUM}$ such that \VTCZbar\
proves
$$
  Y=F_{NUM}(X) \supset \varphi_{NUMONES}(X,Y).
$$
We can define $F_{NUM}$ by the condition
$$
   F_{NUM}(X)(z,y)\lra z\le |X|\wedge y=\numones(z,X).
$$
We can turn this into a proper bit-graph definition
(\ref{e:LFACZ-F}) by using a $\SigZB(\numones)$ formula and
appealing to Lemma \ref{t:open-numones}.
\end{proof}

Unfortunately the analog of the second sentence of Lemma \ref{t:open=sigma}
does not appear to hold.  In general an open formula of \LFTCZ\
is not equivalent to a $\SigZB(\numones)$ formula for the same
reason that a \TCZ\ circuit involving nested threshold gates cannot
be made polynomially equivalent to a circuit with unnested threshold
gates.  Hence we must work harder to prove that \VTCZbar\
is conservative over \VTCZ.

To prove conservativity, we note that  
\VTCZbar\ can be obtained from $\VTCZ(\numones)$ by successively adding
\SigZB-definable functions and their definitions.  This fact together
with Lemma \ref{t:VTCZ-numones} and the following theorem are used to show
both that \VTCZbar\ is conservative
over \VTCZ\ and that all functions in \FTCZ\ are
\SigOneOne-definable in \VTCZ (Corollary \ref{t:VTCZbar-VTCZ}).

\begin{theorem}
\label{t:sig0-comp-new-syms}
Let $\calT$ be an extension of \VZ\ with a
vocabulary $\calL$ which includes the function $\Row$, and suppose that
$\calT$ proves the
defining equation (\ref{e:Row-def}) for $\Row$.  
Suppose that $\calT$ satisfies\\
{\em a)} $\calT$ proves the $\COMP{\SigZB(\calL)}$ scheme, and\\
{\em b)} For each $\SigZB(\calL)$ formula $\alpha$ there is a
\SigOneB\ formula $\beta$ such that $\calT\vdash\alpha\lra\beta$.

Let $\calL'$ extend $\calL$ by adding a function symbol that is
\SigZB-definable from $\calL$ (see Definition~\ref{d:sig0-def}).
Let $\calT'$ be obtained from $\calT$ by adding the function symbol
and its defining axiom.  Then $\calT'$ is conservative over $\calT$,
and a) and  b) hold with $\calT$ replaced by
$\calT'$ and $\calL$ replaced by $\calL'$.
\end{theorem}

\begin{proof}
We will consider the case $\calL' = \calL \cup \{F\}$, where
$F$ is a string function \SigZB\ definable from $\calL$, i.e.,
it has the defining axiom
\begin{align}
\label{e:defining-axiom}
F(\xvec, \Xvec)(u) \lra u < t(\xvec, \Xvec) \wedge \varphi(u, \xvec, \Xvec)
\end{align}
for some $\LTwoA$ term $t$ and $\SigZB(\calL)$ formula $\varphi$.
The case in which $\calL'$ extends $\calL$ by a number function
is handled similarly, except that number variables $w_i$ are used
instead of the string variables $W_i$ in the argument below.

Since $\calT$ proves the $\COMP{\SigZB(\calL)}$ scheme, it follows that it
$\SigOneB(\calL)$-defines $F$.
Therefore $\calT'$ is conservative over $\calT$.

a)
We will show that $\calT'$ proves a slightly modified version of the comprehension axiom
\begin{equation}
\label{e:comp-all-vars}
\exists Z \le \langle \bvec \rangle \forall \zvec < b,\ Z(\zvec) \lra \psi(\zvec)
\end{equation}
for each $\SigZB(\calL')$ formula $\psi$,
where $\zvec$ are all number free variables of $\psi$.
It is straightforward to obtain the usual comprehension axiom scheme
from this.
Also, since $\calT$ extends $\VZ$, it proves this version of $\COMP{\SigZB(\calL)}$.
We will prove \eqref{e:comp-all-vars} by induction on the quantifier depth of $\psi$.

For the base case, $\psi$ is quantifier-free.  Suppose that
$F(\svec_1, \Tvec_1), \ldots, F(\svec_k, \Tvec_k)$
are all occurrences of $F$ in $\psi$.
Note that the terms $\svec_i, \Tvec_i$ may contain $\zvec$ as well as $F$.
Assume further that $\svec_1, \Tvec_1$ do not contain $F$, and
for $ 1 < i \le k$, any occurrence of $F$ in $\svec_i, \Tvec_i$ must be of the form
$F(\svec_j, \Tvec_j)$, for some $j < i$.
We proceed to eliminate $F$ from $\psi$ by using its defining axiom 
(\ref{e:defining-axiom}).

Let $W_1,...,W_k$ be new string variables.
Let $\varphi_1(\zvec, u) \equiv \varphi(u, \svec_1, \Tvec_1)$, and for $2 \le i \le k$,
$\varphi_i(\zvec, u)$ is obtained from $\varphi(u, \svec_i, \Tvec_i)$ 
by replacing every maximal occurrence of any $F(\svec_j, \Tvec_j)$, for $j<i$, by
$W_j^{[\zvec]}$.
Let $t_i$ be obtained from $t(\svec_i, \Tvec_i)$ by the same procedure
(for $i \le k$).   Thus $F$ does not occur in any $\varphi_i$ or $t_i$.
Since $\calT$ proves $\COMP{\SigZB(\calL)}$,
it proves the existence of $W_i$ such that
\begin{align}
\label{e:W-comp}
\forall \zvec < b,\ W_i^{[\zvec]}(u) \lra u < t_i \wedge \varphi_i(\zvec, u)
\qquad
\text{ for $1 \le i \le k$.}
\end{align}
Let $\psi'(\zvec, W_1, \ldots, W_k)$ be obtained from $\psi(\zvec)$ by
replacing each maximal occurrence of $F(\svec_i, \Tvec_i)$
by $W_i^{[\zvec]}$, for $1 \le i \le k$.
Then, by $\COMP{\SigZB(\calL)}$ and the fact that $\calL$ contains $\Row$,
\begin{equation*}
\calT \vdash \exists Z \le \langle \bvec \rangle \forall \zvec < b,\ Z(\zvec) \lra \psi'(\zvec, W_1, \ldots, W_k).
\end{equation*}
Then such $Z$ satisfies $\forall \zvec < b,\ Z(\zvec) \lra \psi(\zvec)$
when each $W_i$ is defined by (\ref{e:W-comp}).

For the induction step, it suffices to consider the case
$\psi(\zvec) \equiv \forall x < t \varphi'(\zvec, x)$.
By the induction hypothesis,
\begin{equation*}
\calT' \vdash \exists Z' \le \langle \bvec, t \rangle \forall \zvec < b \forall x < t,\ Z'(\zvec, x) \lra \varphi'(\zvec, x).
\end{equation*}
Now, by $\COMP{\SigZB}$,
\begin{equation*}
\VZ \vdash \exists Z \le \langle \bvec \rangle \forall \zvec < b,\ Z(\zvec) \lra \forall x < t Z'(\zvec, x).
\end{equation*}

b)
Suppose that
\begin{equation*}
\alpha \equiv \rmQ_1 z_1 < r_1 \ldots \rmQ_n z_n < r_n \psi(\zvec)
\end{equation*}
is a $\SigZB(\calL')$ formula, where $\psi$ is quantifier-free.
Let $\psi'(\zvec, W_1, \ldots, W_k)$ be obtained from $\psi(\zvec)$ as
described  above in the proof of a).  Define
$$
  \alpha'(W_1,...,W_k)\equiv 
      \rmQ_1 z_1 < r_1 \ldots \rmQ_n z_n < r_n \psi'(\zvec,W_1,...,W_k).
$$
For $1\le i\le k$ let $\gamma_i$ be the formula (\ref{e:W-comp}).
Then, $\alpha$ is equivalent in $\calT'$ to
\begin{equation*}
\exists W_1 \le \langle \rvec, t_1 \rangle \ldots
\exists W_k \le \langle \rvec, t_k \rangle,\
(\bigwedge \gamma_i) \wedge \alpha'(W_1, \ldots, W_k).
\end{equation*}
By property b) for $\calT$ we may replace the part of the above formula
following the string quantifier prefix by a \SigOneB\ formula, and thus
we obtain the required \SigOneB\ formula $\beta$ in c) for $\calT'$.
\end{proof}

Let $\VTCZ(\Row, \numones)$ be $\VTCZ$ together with the functions $\Row$
and \numones\ and their defining axioms (\ref{e:Row-def}),
\eqref{e:numones-1}, \eqref{e:numones-2}, \eqref{e:numones-3}.
Since both $\Row$ and \numones\ are \SigOneOne-definable in \VTCZ\
it follows that $\VTCZ(\Row, \numones)$ is conservative over \VTCZ.

\begin{lemma}
\label{t:VTCZ-numones}
Let $\calT$ be the theory $\VTCZ(\Row, \numones)$.  Then $\calT$
satisfies hypotheses a) and b) in Theorem \ref{t:sig0-comp-new-syms}.
\end{lemma}

\begin{proof}
First note that $\VTCZ(\Row)$ satisfies a) and b) by Lemma \ref{l:row-pairing}.
We will prove the present
lemma by modifying the proof of Theorem \ref{t:sig0-comp-new-syms}
applied as if $\calT'$ is\\ $\VTCZ(\Row,\numones)$ and $\calT$ is $\VTCZ(\Row)$.

Proceeding as in the proof of a), we want to show that $\calT'$
proves (\ref{e:comp-all-vars})
where $\psi(\zvec)$ is a $\SigZB(\Row, \numones)$ formula.  Arguing as
before, it suffices to consider the base case of the induction, where
$\psi$ is quantifier-free, and \numones\ plays the role of $F$ in the
previous argument.  Thus $\numones(s_1,T_1),...,\numones(s_k,T_k)$ are
all occurrences of \numones\ in $\psi$, ordered as before.  We proceed
to eliminate the occurrences of \numones\ from $\psi$ using 
(\ref{e:numones-NUM}).

Let $w_1,...,w_k$ be new number variables.  Let $s'_1\equiv s_1$,
and for $2\le i\le k$ let $s'_i$ be obtained from $s_i$ by replacing
every maximal occurrence of $\numones(s_j,T_j)$, for $j<i$, by $w_j$.
Let $T'_i$ be obtained from $T_i$ in the same way. 
(Thus \numones\ does not occur in any of the $s_i$'s and $T_i$'s.)
Let $Y_1,...,Y_k$ be new string variables.
By Claim~\ref{t:VTCZ-boot} (below), for $1 \le i \le k$,
$\VTCZ(\Row)$ proves the existence of $Y_i$ such that
\begin{equation}
\label{e:VTCZ-Row}
   \forall \zvec <b \, \forall w_1\le s'_1...\forall w_k\le s'_k \ 
           \varphi_\NUMONES(T'_i, Y^{[\zvec,\wvec]}_i).
\end{equation}
If $Y_1,...,Y_k$ each satisfies (\ref{e:VTCZ-Row}), and $w_1,...,w_k$
each satisfies $Y^{[\zvec,\wvec]}_i(s'_i,w_i)$ then
by (\ref{e:numones-NUM}) each $w_i$ must have its intended
value $\numones(s_i,T_i)$.  Thus $\VTCZ(\Row,\numones)$ proves
$$
   \forall \zvec <b \forall \wvec \le \vec{s'}, \
    (\bigwedge_{i=1}^k (\varphi_\NUMONES(T'_i,Y^{[\zvec,\wvec]}_i)
      \wedge Y^{[\zvec,\wvec]}_i(s'_i,w_i)) \ \supset \
    \bigwedge_{i=1}^k w_i = \numones(s_i,T_i)
$$
Let $\psi'(\zvec,w_1,...,w_k)$ be obtained from $\psi(\zvec)$ by
replacing each maximal occurrence of\\ $\numones(s_i,T_i)$ by $w_i$,
for $1\le i \le k$.  Then by $\COMP{\SigZB(\Row)}$ we have $\VZ(\Row)$ proves
$$
\exists Z \le \langle \bvec \rangle \forall \zvec < b,\
Z(\zvec) \lra \exists w_1\le s'_1...\exists w_k\le s'_k,\
 (\bigwedge Y_i^{[\zvec,\wvec]}(s'_i,w_i))\wedge \psi'(\zvec, w_1, \ldots, w_k).
$$
Then such $Z$ satisfies $\forall \zvec < b,\ Z(\zvec)\lra \psi(\zvec)$
when each $Y_i$ satisfies (\ref{e:VTCZ-Row}).

To prove b), suppose that 
\begin{equation*}
\alpha \equiv \rmQ_1 z_1 < r_1 \ldots \rmQ_n z_n < r_n \psi(\zvec)
\end{equation*}
is a $\SigZB(\Row,\numones)$ formula, where $\psi$ is quantifier-free.
Then, using the notation of the proof of a) above, 
$\VTCZ(\Row, \numones)$ proves
\begin{multline*}
   \alpha \lra \exists \Yvec \le \tvec, \\
            (\forall \zvec < b \forall \wvec \le \vec{s'}
        \bigwedge \varphi_\NUMONES(T'_i,Y_i^{[\zvec.\wvec]})) \ \wedge \
 \Qvec \zvec < \rvec \exists \wvec \le \vec{s'}, \
        \psi'(\zvec,\wvec) \wedge \bigwedge Y_i^{[\zvec,\wvec]}(s'_i,w_i)
\end{multline*}
for suitable terms $\tvec$ bounding $\Yvec$.
The RHS is a $\SigOneB(\Row)$ formula which, by Lemma \ref{l:row-pairing}
is equivalent to a \SigOneB\ formula.
\end{proof}

To complete the proof of Lemma~\ref{t:VTCZ-numones}, we show that
for $1 \le i \le k$, \VTCZ\ proves the existence of $Y_i$ which satisfies
\eqref{e:VTCZ-Row}.
It suffices to show that $\VTCZ(\Row)$ proves the existence of multiple
``counting arrays'' for polynomially many strings.

\begin{claim}
\label{t:VTCZ-boot}
The theory $\VTCZ(Row)$ proves the existence of $Y$ such that
$$\forall u < b\ \varphi_\NUMONES(X^{[u]}, Y^{[u]})$$
\end{claim}

\begin{proof}
We construct (using $\COMP{\SigZB(Row)}$) multiple counting arrays $Y^{[0]}, \ldots, Y^{[b-1]}$
from the counting array $Y'$ for a ``big'' string $X'$, which
is obtained from the strings $X^{[0]}, \ldots, X^{[b-1]}$ simply by
concatenating them.
More precisely, let $X'$ be defined by
$$X'(u|X| + x) \lra X^{[u]}(x),\ \ \ \mbox{for $x < |X|, u < b$.}$$
Thus $X'(u|X|), \ldots, X'((u+1)|X| - 1)$ is a copy of $X^{[u]}$.
Therefore
$$\numones(z, X^{[u]}) = \numones(u|X| + z, X') - \numones(u|X|, X').$$

Let $Y'$ be the counting array for $X'$, i.e.,
$Y'(z, y) \Lra \numones(z, X') = y$.
Then $Y^{[u]}(z, y) \Lra y = \numones(u|X| + z, X') - \numones(u|X|, X')$.
Hence
$$Y^{[u]}(z, y) \lra \exists y_1, y_2 \le |X'|,\ Y'(u|X|, y_1) \wedge
Y'(u|X|+z, y_2) \wedge y + y_1 = y_2$$
\end{proof}

\begin{corollary}
\label{t:VTCZbar-VTCZ}
\VTCZbar\ is a conservative extension of \VTCZ.  Every function
in \LFTCZ\ is \SigOneOne-definable in \VTCZ.
\end{corollary}

\begin{proof}
$\VTCZ(Row,\numones)$ is conservative over \VTCZ\ because $Row$
and \numones\ are \SigOneB-definable in \VTCZ.
According to Lemma \ref{t:VTCZ-numones} and
Theorem \ref{t:sig0-comp-new-syms}, \VTCZbar\ is the union of a sequence of
conservative extension of $\VTCZ(Row,\numones)$
satisfying a) and b).
Thus (by compactness) \VTCZbar\ is conservative over \VTCZ. 
Each of these extensions
is obtained by adding a $\SigZB(\calL)$-definable
function.  The graph of each such function has a $\SigZB(\calL)$ definition,
which by b) is provably equivalent to a \SigOneB\ formula (in the
language of \VTCZ).  Hence this function is \SigOneB-definable in
\VTCZbar\ and hence in \VTCZ.
\end{proof}

The above corollary proves one direction of Theorem \ref{t:definable-fcns}
for the case of \VTCZ.  
For the other direction we need witnessing theorems, which are
the subject of Subsection \ref{s:wit-thms}.

Recall that each string function $F \in \FTCZ$ has a defining axiom
according to our construction of \LFTCZ\
(see Definition~\ref{d:LFTCZ} and Lemma~\ref{t:LFTCZ-repre}).
In fact, there is a finite sequence of $\FTCZ$ functions $F_1, \ldots, F_n$
that are involved in defining $F$.
Let $\calL(F)$ denote this sequence of functions (including $F$),
and let $\AX{\calL(F)}$ be the set of their defining axioms.
The following corollary is proved similarly to Corollary~\ref{t:VTCZbar-VTCZ}.

\begin{corollary}
\label{t:VTCZ-F}
For each $F \in \FTCZ$, \VTCZbar\ is a conservative extension of
the theory $\VTCZ \cup \AX{\calL(F)}$.
\end{corollary}

\ignore{
\subsubsection{The Theories $\VZbar(m)$ And \VACCbar}

For each $m \ge 2, m \in \N$,
the function $\modulo_m(X, z)$ can be defined as follows.
\begin{align}
\label{e:modulo-1}
\modulo_m(X, 0) = 0 \\
\label{e:modulo-2}
X(z) \wedge \modulo_m(X, z) = m-1 \supset \modulo_m(X, z + 1) = 0 \\
\label{e:modulo-3}
X(z) \wedge \modulo_m(X, z) < m-1 \supset \modulo_m(X, z + 1) = \modulo_m(X, z)
+ 1 \\
\label{e:modulo-4}
\neg X(z) \supset \modulo_m(X, z+1) = \modulo_m(X, z)
\end{align}

\begin{definition}
\label{d:LFACZm}
For each $m \in \N, m \ge 2$, the vocabulary $\LFACZm$
is defined similarly to \LFTCZ\ (Definition~\ref{d:LFTCZ}),
except for $\numones$ is replaced by $\modulo_m$.
\end{definition}

\begin{definition}
\label{d:VZmbar}
For each $m \in \N, m \ge 2$, $\VZbar(m)$ is the theory
over $\LFACZm$, whose axioms
are the universal closure of {\bf B1 - B11}, {\bf B12$'$}, {\bf B12$''$},
{\bf L1}, {\bf L2}, {\bf SE$'$}, {\bf SE$''$},
the defining axioms \eqref{e:fSE-1}, \eqref{e:fSE-2}, \eqref{e:fSE-3} for
$\fSE$,
the defining axioms \eqref{e:modulo-1}, \eqref{e:modulo-2}, \eqref{e:modulo-3},
\eqref{e:modulo-3} for $\modulo_m$,
and the defining axioms
\eqref{e:LFACZ-F} and \eqref{e:LFACZ-f-1}, \eqref{e:LFACZ-f-2},
\eqref{e:LFACZ-f-3}
for the functions of $\LFACZm$.
Let $\VACCbar = \bigcup_{m \ge 2}\VZbar(m)$.
\end{definition}

We summarize some results for these theories, which can be proved
in the same way as we have done for \VTCZbar.

\begin{theorem}
For each $m \in \N, m \ge 2$, $\VZbar(m)$ is a conservative
extension of $\VZ(m)$.
The open formulas of $\LFACZm$ represent precisely the class $\ACZ(m)$,
and the functions of $\LFACZm$ represent precisely $\FACZ(m)$.
Every function in $\FACZ(m)$ is $\SigOneOne$ definable in $\VZ(m)$.
Similarly for \VACCbar\  with respect to \VACC\ and $\FACC$.
\end{theorem}
}

\subsection{Witnessing Theorems}
\label{s:wit-thms}

In this subsection we will prove the remaining direction of
Theorem \ref{t:definable-fcns}, namely that the
\SigOneOne-definable functions in each of our various theories are in
the appropriate complexity class.

We will use the proof system $\LKTwo$ \cite{Cook:02:notes} which extends
$\LK$ (see e.g. \cite{Buss:98:book}) by the introduction rules for
string variable quantifiers.
It is convenient to distinguish between {\em bound variables}
(which are denoted by $x, y, z, \ldots$ for number variables,
and $X, Y, Z, \ldots$ for the string variables)
and {\em free variables}
(which are denoted by $a, b, c, \ldots$ for number variables,
and $\alpha, \beta, \gamma, \ldots$ for the string variables).
Recall the definition of $\LFACZ$ in Definition~\ref{d:LFACZ}.

\begin{theorem}
[General Witnessing Theorem]
\label{t:witnessing}
Suppose that $\calL$ extends $\LFACZ$ and that it satisfies
conditions {\em b, c} in Definition~\ref{d:LFACZ} with $\LFACZ$ replaced by $\calL$.
Suppose that $\calT$ is an open theory extending $\VZbar$ 
and that $\calT$ contains the defining axiom \eqref{e:LFACZ-F}
for each function $F_{\alpha,t}$ of $\calL$,
and the defining axioms
\eqref{e:LFACZ-f-1}, \eqref{e:LFACZ-f-2} and \eqref{e:LFACZ-f-3}
for each function $f_{\alpha,t}$ of $\calL$.
Then for each theorem $\exists \Zvec \varphi(\avec, \alphavec, \Zvec)$
of $\calT$, where $\varphi$ is a $\gSigOneB$ formula,
there are functions $\Fvec$ of $\calL$ such that
$$
\calT \vdash \forall \xvec \forall \Xvec\ \varphi(\xvec, \Xvec, \Fvec(\xvec, \Xvec)).
$$
\end{theorem}

\begin{proof}
Note that when $\varphi$ is an open formula, the Theorem is an application of
Herbrand Theorem.
To prove the current Theorem for the general case
we will follow the proof theoretic approach
and examine the $\LKTwoT{\calT}$ proofs (i.e., proofs in $\LKTwo$
with non-logical axioms from \calT).
In particular, we will explicitly witness the string existential
quantifiers in every line of an {\em anchored}
\cite{Buss:98:book,Cook:02:notes}
(also known as a {\em free-cut free}) proof of
$\exists \Zvec \varphi(\avec, \alphavec, \Zvec)$
by functions from $\calL$.
This is explained below.
First, the following Claim will simplify our arguments.

{\bf Claim}: For each $\SigZB(\calL)$ formula $\phi(\xvec, \Xvec)$,
there is an open formula $\psi(\xvec, \Xvec)$ of $\calL$ such that
\[\calT \vdash \psi(\xvec, \Xvec) \lra \phi(\xvec, \Xvec).\]

Note that on page~\pageref{d:fSE} we have used $\fSE$ to eliminate
an implicit quantifier $\exists z < |X|$ in the axiom {\bf SE}.
The proof of this Claim is similar and is omitted.

Now for simplicity, 
assume that we are to witness a single variable $Z$ 
in $\exists Z \varphi(\xvec, \Xvec, Z)$,
where $\varphi$ is a \gSigOneB\ formula in prenex form.
Consider the most interesting case when $\varphi$ is of the form:
\begin{equation}\label{e:witness-0}
\varphi(\avec, \alphavec, Z) \equiv
\forall y_n < b\ \exists Y_n < b \ \ldots\ \forall y_1 < b \ \exists Y_1 < b\
\theta(\avec, \alphavec, \yvec, \Yvec, Z)
\end{equation}
By the above Claim we can assume that $\theta$ is an open formula of $\calL$.

An {\em anchored} $\LKTwoT{\calT}$ proof
$\pi$ is a proof
in the system $\LKTwo$ with additional non-logical axioms
the instances of axioms of \calT,
and the cut formulas of $\pi$ are restricted to these instances only.
By a standard argument,
there exists an anchored $\LKTwoT{\calT}$ proof $\pi$ of
$\exists Z\ \varphi(\avec, \alphavec, Z)$.
Since $\calT$ is an open theory, the cut formulas in $\pi$ are quantifier-free.
Thus quantified formulas in $\pi$ can appear only in the succedents,
and must be of one of the two forms below
(we will not mention the bound $b$ on variables)
\begin{align}
\label{e:witness-1}
\exists Y_m \ \forall y_{m-1} \ldots\ \forall y_1 \ \exists Y_1\
\theta(\avec, \alphavec, c_n, \ldots, c_{m}, y_{m-1}, \ldots, y_1, T_n, \ldots, T_{m+1}, Y_{m}, \ldots, Y_1, T)\\
\label{e:witness-2}
\forall y_m\ \exists Y_m \ \ldots\ \forall y_1 \ \exists Y_1\
\theta(\avec, \alphavec, c_n, \ldots, c_{m+1}, y_m, \ldots, y_1, T_n, \ldots, T_{m+1}, Y_{m}, \ldots, Y_1, T)
\end{align}
($c_i$'s are free number variables, and $T$, $T_j$'s are string terms
which do not involve bound variables $y_i$'s and $Y_j$'s).
Therefore the only quantifier introduction rules can be used in $\pi$ are the number $\rright{\forall}$ rule
and the string $\rright{\exists}$ rule.
Also, the $\wedge$, $\vee$ and $\neg$ introduction rules can only
be applied to quantifier-free formulas.

We will prove by induction on the length of $\pi$ that
for each sequent $\calS$ of $\pi$, there are functions $F_i$'s of $\calL$
(called the {\em witnessing functions} of $\calS$)
so that the sequent $\calS'$, which is constructed from $\calS$ and $F_i$'s
as described shortly,
is a theorem of $\calT$.
Essentially $F_i$'s are the witnessing functions that compute the existentially
quantified string variables of $\calS$,
and $\calS'$ is constructed from $\calS$ by explicitly mentioning these witnessing functions.
Suppose that $\calS = \Lambda \longra \Gamma$
(note that $\Lambda$ contains only open formulas),
then $\calS' = \Lambda \longra \Gamma'$,
where $\Gamma'$ consists of the following (quantifier-free) formulas.
(We drop mention of $\avec, \alphavec$ in $\theta$ as well as in $F_i$'s.
Note that the functions $F_i$'s may contain free variables that are present in $\calS$.
We write $\cvec_{[i,k]}$ for $c_i, \ldots, c_k$,
and similarly for $\bvec_{[i,k]}$ and $\Tvec_{[i,k]}$.)
\begin{itemize}
\item
All open formulas in $\Gamma$
\item
For each formula of the form \eqref{e:witness-1} in $\Gamma$,
the formula
\begin{equation}
\label{e:witness-1'}
\theta(\cvec_{[n,m]}, \bvec_{[m-1,1]}, \Tvec_{[n,m+1]},
F_{m}(\cvec_{[n,m]}), \ldots, 
F_1(\cvec_{[n,m]}, \bvec_{[m-1,1]}), T)
\end{equation}
\item
For each formula of the form \eqref{e:witness-2} in $\Gamma$,
the formula
\begin{equation}
\label{e:witness-2'}
\theta(\cvec_{[n,m+1]}, \bvec_{[m,1]}, \Tvec_{[n,m+1]},
F_{m}(\cvec_{[n,m+1]}, b_m), \ldots, 
F_1(\cvec_{[n,m+1]}, \bvec_{[m,1]}), T)
\end{equation}
\end{itemize}
(In \eqref{e:witness-1'} and \eqref{e:witness-2'},
the free variables $b_i$'s do not appear anywhere else in $\calS'$.)

The base case holds trivially, since the axioms of $\calT$ are open formulas.
For the induction step, we consider the inference rules that might be used in $\pi$.

{\bf Case I} (String $\rright{\exists}$):
Suppose that $\mathcal{S}$ is the bottom sequent of the inference
\begin{align*}
\begin{prooftree}
\calS_1
\justifies
\calS
\end{prooftree}
=
\begin{prooftree}
\Lambda\longrightarrow\Gamma, \psi(T_{m+1})
\justifies
\Lambda\longrightarrow\Gamma, \exists Y_{m+1}\psi(Y_{m+1})
\end{prooftree}
\end{align*}
where $\psi$ is as \eqref{e:witness-2}.
By the induction hypothesis, $\calS_1'$ is a theorem of $\calT$.
We obtain $\calS'$ from $\calS_1'$ by taking $F_{m+1}$ to be
the function defined by $T_{m+1}$.

{\bf Case II} (Number $\rright{\forall}$):
Note that this rule can be applied to only formulas of the form \eqref{e:witness-1}.
Also in this case, the free variable $c_m$ must not appear anywhere else in $\calS$.
In the witnessing functions that occur in \eqref{e:witness-1'}, $c_m$ is replaced by $b_m$.
No new function is required.

{\bf Case III} (Cut):
Note that the cut formula is an open formula.
Suppose that $\calS$ is derived from $\calS_1$ and $\calS_2$ using
the cut rule:
\begin{align*}
\begin{prooftree}
\calS_1 \qquad \calS_2
\justifies
\calS
\end{prooftree}
=
\begin{prooftree}
\Lambda \longra \Gamma, \psi
\qquad
\Lambda, \psi \longra \Gamma
\justifies
\Lambda \longra \Gamma
\end{prooftree}
\end{align*}
where $\psi$ is an open formula of $\calL$.
The witnessing functions of $\calS$ is defined from the witnessing
functions of $\calS_1$ and $\calS_2$ as follows:
$$
F_i(z) \lra (\neg \psi \wedge F^1_i(z))\ \vee \ (\psi \wedge F^2_i(z))
$$

{\bf Case IV} (Weakening rule):
If $\calS$ is obtained from $\calS_1$ by the weakening rule,
then $\calS'$ can be obtained from $\calS_1'$ by the same rule.
When the additional formula in $\calS$ is a $\gSigOneB$
formula (of the form \eqref{e:witness-1} or \eqref{e:witness-2}),
the witnessing functions can be the constant string
function $\setZ$.

{\bf Case V} (Contraction rule):
Suppose that $\calS$ is derived from $\calS_1$ using the contraction rule.
Consider the interesting case where the formula removed from $\calS_1$ is
a $\gSigOneB$ formula.
\begin{align*}
\begin{prooftree}
\calS_1
\justifies
\calS
\end{prooftree}
=
\begin{prooftree}
\Lambda \longra \Gamma, \psi(\avec, \alphavec, \cvec), \psi(\avec, \alphavec, \cvec)
\justifies
\Lambda \longra \Gamma, \psi(\avec, \alphavec, \cvec)
\end{prooftree}
\end{align*}
($\psi(\avec, \alphavec, \cvec)$ is of the form \eqref{e:witness-1} or \eqref{e:witness-2}).
Note that the two occurrences of $\psi$ in $\calS_1$ may have
different collections of witnessing functions in $\calS_1'$.
However, if $\forall \zvec \psi(\avec, \alphavec, \zvec)$ is to be true,
then at least one of the two collections is correct.
The witnessing functions in $\calS'$ are defined using this information.

Formally, consider the case of \eqref{e:witness-1}, and assume that
corresponding to the two occurrences of $\psi(\avec, \alphavec, \cvec)$ in
$\calS_1$, we have the following formulas in $\calS_1'$
(see \eqref{e:witness-1'}):
\begin{align*}
\theta^1(\cvec, \bvec) \equiv \theta(\cvec, \bvec, \Tvec, F_m^1(\cvec), \ldots, F_1^1(\cvec, \bvec))\\
\theta^2(\cvec, \bvec) \equiv \theta(\cvec, \bvec, \Tvec, F_m^2(\cvec), \ldots, F_1^2(\cvec, \bvec))
\end{align*}
In general, the witnessing functions of $\calS'$ are
$$
F_i(\cvec, \bvec)(x) \lra
(\forall \zvec \ \forall \yvec\ \theta^1(\zvec, \yvec) \wedge F_i^1(\cvec, \bvec)(x))\ \vee \
(\neg \forall \zvec \ \forall \yvec\ \theta^1(\zvec, \yvec) \wedge F_i^2(\cvec, \bvec)(x))
$$

{\bf Case VI} (Other rules):
When $\varphi$ is in prenex form \eqref{e:witness-0}, the introduction rules for $\wedge, \vee, \neg$ can
be applied to only quantifier-free formulas.
No new function is required.
In general, handling these rules is more complicated, but is straightforward.
Similarly, if $\calS$ is obtained from $\calS_1$ by
the exchange rule, then $\calS'$ can be derived from $\calS_1'$
by the same rule.
\end{proof}

\begin{corollary}
[{Witnessing Theorems for \VTCZ}] 
\label{t:VTCZ-wit}
For each theorem $\exists \Zvec \varphi(\xvec, \Xvec, \Zvec)$
of $\VTCZ$ where $\varphi$ is $\gSigOneB$,
there are string functions $\Fvec \in \FTCZ$, such that
\[\VTCZ \cup \AX{\calL(\Fvec)} \vdash
\varphi(\xvec, \Xvec, \Fvec(\xvec, \Xvec)).\]
\end{corollary}

\begin{proof}
Suppose that $\VTCZ \vdash \exists \Zvec \varphi(\xvec, \Xvec, \Zvec)$,
so $\VTCZbar \vdash \exists \Zvec \varphi(\xvec, \Xvec, \Zvec)$.
By Theorem~\ref{t:witnessing}, there are string functions $\Fvec \in \LFTCZ$
such that $\VTCZbar \vdash \varphi(\xvec, \Xvec, \Fvec(\xvec, \Xvec))$.
The conclusion follows from Corollary~\ref{t:VTCZ-F}.
\end{proof}

Note that similar witnessing theorems hold for the universal theory
\VTCZbar. 

\begin{corollary}
[{The remaining direction of Theorem \ref{t:definable-fcns}}]
The \SigOneOne-definable function in \VTCZ\ 
are in \FTCZ. 
\end{corollary}

\subsection{Theories for Other Subclasses of \Ptime}

In this subsection, we will apply Theorem~\ref{t:sig0-comp-new-syms}
and Theorem~\ref{t:witnessing} to develop finitely axiomatizable theories for other
uniform subclasses of \Ptime\ in the same style as \VTCZ.
Let $F$ be a polynomial time string function and let $\boldC$ be
the class of two-sorted relations which are
\ACZ-reducible to $F$.
Note that the associated function class $\FC$ is usually defined using
the graphs (for number functions) and bit graphs (for string functions)
(page \pageref{d:FC}).
But it can be equivalently defined as the class of
functions which are \ACZ-reducible to $F$.
In the case of \TCZ, $F$ is essentially the string function computing
the ``counting array'', whose graph is given by \NUMONES.

We add to \VZ\ a $\SigOneB$ axiom $\AXIOM_F$ which 
formalizes the polytime algorithm that computes $F$. 
(Thus $\AXIOM_F$ is a generalization of \NUMONES.)
We will show that the resulting theory $\calT$ characterizes $\boldC$.
The proof is almost identical to the proof in the case of \VTCZ:
we show that the universal theory \calTbar\
(obtained from $\LTwoA(\Row, F)$ in the same way that \VTCZbar\ is obtained from $\LTwoA(\Row, \numones)$)
characterizes the same class.
The main task is to prove the analogue of Lemma~\ref{t:VTCZ-numones},
i.e., $\calT(\Row, F)$ satisfies hypotheses a) and b) in Theorem~\ref{t:sig0-comp-new-syms}.
Our choice of the axiom $\AXIOM_F$ will make this step readily obtainable.
\footnote{Note the fact that $\SigOneOne$ theorems of $\calT$ can be
witnessed by functions of $\FC$ follows easily from Herbrand's Theorem.
The General Witnessing Theorem offers more than we need here.
It is necessary in Section~\ref{s:RSUV} where we show that \VTCZ\
is RSUV isomorphic to \DelCR.}

The universal defining axiom for $F$ is obtained from a modified version
of Cobham's recursion theoretic characterization of the polytime functions.
Here we use the fact that each polytime function can be obtained from
\ACZ\ functions by composition and at most one application of the
{\em bounded recursion} operation.
In each complexity class of interest it turns out that a suitable
function $F$ complete for the class can be defined by such a recursion
of the form
\begin{align*}
F(0, X) = (\Init(X))^{<t(0, |X|)}\\
F(x+1, X) = (\Next(x, X, F(x, X)))^{<t(x+1, |X|)}
\end{align*}
where $\Init(X)$ and $\Next(x, X, Y)$ are \ACZ\ functions, $t(x,y)$
is a polynomial, and $X^{<y}$ is the initial segment of $X$ of length $y$.

We will first define the universal theory $\calTbar$ in the same manner that
we have defined \VTCZbar.
Here we will not introduce the new functions $\Init, \Next$ and $X^{<y}$
but will use their $\SigZB$ definitions instead.
In other words, $F$ can be defined as follows:
\begin{align}
\label{e:define-F-1}
F(0, X)(z) \lra  z < t(0, |X|) \wedge \varphi_\Init(z, X)\\
\label{e:define-F-2}
F(x + 1, X)(z) \lra z < t(x+1, |X|) \wedge  \varphi_\Next(z, x, X, F(x, X))
\end{align}
where $\varphi_\Init$ and $\varphi_\Next$ are the $\SigZB$ bit definitions
of $\Init$ and $\Next$ respectively.
(Note that in \VZbar, $\varphi_\Init$ and $\varphi_\Next$ are equivalent to
open formulas of $\LFACZ$.)

The language \LFC\ of functions in \FC\ is defined in the same way as \LFTCZ\
(Definition~\ref{d:LFTCZ}),
except for $\numones$ is replaced by $F$.
The theory $\calTbar$ is defined similarly to \VTCZbar\ (Definition~\ref{d:VTCZbar}),
with the defining axioms \eqref{e:define-F-1}
and \eqref{e:define-F-2} of $F$ replacing the defining axioms of $\numones$.
The following Corollary follows from Theorem~\ref{t:witnessing}.
\begin{corollary}
\label{t:VCbar-witness}
For each theorem $\exists \Zvec \varphi(\avec, \alphavec, \Zvec)$
of $\calTbar$, where $\varphi$ is a $\gSigOneB$ formula,
there are functions $\Fvec$ of $\LFC$ such that
$$
\calTbar \vdash \forall \xvec \forall \Xvec\ \varphi(\xvec, \Xvec, \Fvec(\xvec, \Xvec)).
$$
\end{corollary}
On the other hand, $\calTbar$ proves the $\COMP{\SigZB(\LFC)}$ scheme,
and thus can $\SigOneB(\LFC)$-define all functions of \LFC.

Now we will define $\calT$.
Our choice for the $\SigOneB$ defining axiom of $F$ comes from the above definition of $F$
given in \eqref{e:define-F-1}, \eqref{e:define-F-2}.
In order to prove Claim \ref{t:VC-F} (see the discussion below)
we will not compute $F$
for a single value of $X$, but rather multiple (i.e., polynomially many) values of $X$.
Let $\varphi_F(a, b, X, Y)$ be the formula stating that
$Y$ encodes simultaneously the $b$ recursive computations of $F(a, X^{[0]}), \ldots, F(a, X^{[b-1]})$
using the definition of $F$.
More precisely, let $\varphi_F(a, b, X, Y)$ be
\begin{equation}
\label{e:varphi-F}
\begin{split}
\forall y < b,\ & [
\forall z < a Y^{[y, 0]}(z) \lra \varphi_\Init(z, X^{[y]}) \ \wedge\\
& \forall x < a \forall z < a,\ Y^{[y, x+1]}(z) \lra z < t(x+1, |X^{[y]}|) \wedge \varphi_\Next(z, x, X^{[y]}, Y^{[y, x]})]
\end{split}
\end{equation}

\begin{definition}
Let $\AXIOM_F$ be $\forall a \forall b \forall X \exists Y \le \langle b, t(a, |X|) \rangle \varphi_F(a, b, X, Y)$.
The theory \calT\ is $\VZ$ extended by the axiom $\AXIOM_F$.
\end{definition}

Since \VZ\ is finitely axiomatizable, so is $\calT$.

\begin{lemma}
\label{t:calT-calTbar}
\calTbar\ is a conservative extension of \calT\
and satisfies the hypotheses a), b) of Theorem~\ref{t:sig0-comp-new-syms}.
\end{lemma}

\begin{proof}
The proof is the same as the first part of the proof of Corollary~\ref{t:VTCZbar-VTCZ}.
First, let $\calT(\Row, F)$ be \calT\ together with the functions $\Row$ and $F$
and their defining axioms \eqref{e:Row-def}, \eqref{e:define-F-1}, \eqref{e:define-F-2}.  Note that $\calT(\Row, F)$ is conservative over $\calT$,
because both $\Row$ and $F$ are \SigOneB-definable in \calT.
(The \SigOneB-definability of $F$ follows by $\AXIOM_F$.)
Assume that
\begin{claim}
\label{t:VC-F}
$\calT(\Row, F)$ satisfies hypotheses a) and b) in
Theorem~\ref{t:sig0-comp-new-syms}.
\end{claim}
Then $\calTbar$ is obtained from $\calT(\Row, F)$ by a series of conservative
extensions satisfying hypotheses a) and b) of Theorem~\ref{t:sig0-comp-new-syms}.

It remains to prove the Claim.
We proceed as in the proof of Theorem~\ref{t:sig0-comp-new-syms}.
In fact, it suffices to show that $\calT(\Row, F)$ proves the existence
of $W$ such that for all $\zvec < b$, $W^{[\zvec]}$ is the intended value of $F(s, T)$
where $s, T$ are terms of $\LTwoA(\Row)$ which may contain $\zvec$.
(Compare to \eqref{e:W-comp}.)
Using $\AXIOM_F$, such $W$ can be constructed using $\COMP{\SigZB}$.
\end{proof}

\begin{corollary}
\label{t:VC-define-funcs}
The $\SigOneOne$-definable (and the $\SigOneB$-definable) functions in \calT\
are precisely those in \LFC.
\end{corollary}

\begin{proof}
Each function of $\LFC$ has a $\SigZB(\LFC)$ definition, which is
equivalent in \calTbar\ to a \SigOneB\ formula, by Lemma~\ref{t:calT-calTbar}.
It is therefore $\SigOneB$ definable in \calT.

On the other hand, $\SigOneOne$ theorems of \calTbar\ (and hence of \calT)
are witnessed by \FC\ functions, as shown in Corollary~\ref{t:VCbar-witness}.
\end{proof}

\ignore{
\begin{lemma}
\label{t:VC-F}
Let $\calT(\Row, F)$ be \calT\ together with the functions $\Row$ and $F$
and their defining axioms \eqref{e:Row-def}, \eqref{e:define-F-1}, \eqref{e:define-F-2}.
Then
$\calT(\Row, F)$ satisfies hypotheses a) and b) in Theorem~\ref{t:sig0-comp-new-syms}.
\end{lemma}

\begin{proof}
We proceed as in the proof of Theorem~\ref{t:sig0-comp-new-syms}.
In fact, it suffices to show that $\calT(\Row, F)$ proves the existence
of $W$ such that for all $\zvec < b$, $W^{[\zvec]}$ is the intended value of $F(s, T)$
where $s, T$ are terms of $\LTwoA(\Row)$ which may contain $\zvec$.
(Compare to \eqref{e:W-comp}.)
Using $\AXIOM_F$, such $W$ can be constructed using $\COMP{\SigZB}$.
\end{proof}
}

Note that the axiom \NUMONES\ is a special case of $\AXIOM_F$.
It is ``nicer'' than $\AXIOM_F$ in the sense that it encodes
only a single computation of $\numones$.
In fact, Claim~\ref{t:VTCZ-boot} shows that $\VTCZ(\Row)$ proves $\AXIOM_\numones$.
We need this in order to show that \VTCZ\ satisfies the hypotheses a) and b)
of Theorem~\ref{t:sig0-comp-new-syms} (Lemma~\ref{t:VTCZ-numones}).
However, the proof of this Claim is rather {\em ad hoc}.

In general,
our choice of $\AXIOM_F$ guarantees that $\calT(\Row, F)$ satisfies the hypotheses a) and b)
of Theorem~\ref{t:sig0-comp-new-syms}
(as shown in Claim~\ref{t:VC-F}).
Thus to go further and obtain ``nicer'' axiom than $\AXIOM_F$
(in the style of \NUMONES), it remains to prove the analogue of Claim~\ref{t:VTCZ-boot}.
These proofs may differ for different chosen functions $F$.
Some examples are given below.

\ignore{
For each particular class, if one wishes to develop an associated theory
in the style of \VTCZ\ (i.e., the additional axiom encodes a single computation
rather multiple computations as in $\AXIOM_F$),
then our result 
We will give as example some other theories, in each case an analogue
of Claim~\ref{t:VTCZ-boot} is necessary to obtain the equivalence of
$\AXIOM_F$ in the style of \NUMONES.
Next, we will give some examples of using Corollary~\ref{t:VC-define-funcs}
to obtain theories characterizing a number of well-known classes.
}

\subsubsection{Theories for $\ACZ(m)$ and \ACC}

Theories for the complexity classes $\ACZ(m)$ and $\ACC$ are defined
in the same way that \VTCZ\ is defined for the class \TCZ.
For $m \ge 2$, $\varphi_{\MOD_m}(X, Y)$ is the formula stating that $Y$ is
the ``counting modulo $m$'' array for $X$:
\begin{equation}
\label{e:MOD-m}
\begin{split}
\varphi_{\MOD_m}(X,Y)\equiv &
[\forall z\le |X| \exists! y < m Y(z,y)]\ \wedge\ 
Y(0,0)\ \wedge\
\forall z<|X| \forall y < m,\\
& Y(z,y)\supset [(X(z)\supset Y(z+1, \ y+1 \mod{m})) \wedge
(\neg X(z)\supset Y(z+1,y))].
\end{split}
\end{equation}
Here, we identify the natural number $m$ with the corresponding numeral $\numm$.
We take $\varphi(y \mod{m})$ as an abbreviation for 
\begin{equation}
\label{e:mod-m}
\exists r < m, \exists q \le y,\ y = qm + r \wedge \varphi(r).
\end{equation}
Thus if $\varphi(y)$ is \SigZB, then $\varphi(y\mod{m})$ is also \SigZB.

\begin{definition}
For each $m\ge 2$, let 
$\MOD_m \equiv \forall X \exists Y \varphi_{\MOD_m}(X,Y)$.  Then
\begin{eqnarray*}
\VZ(m) & = & \VZ \cup \{\MOD_m\}  \\
\VACC & = & \VZ \cup \{\MOD_m \mid m \ge 2 \}.
\end{eqnarray*}
\end{definition}
Note that the string $Y$ in $\MOD_m$ can be bounded by
$\langle |X|, m \rangle$.

The following Theorem can be proved in the same way as Theorem~\ref{t:definable-fcns}:
\begin{theorem}
The $\Sigma_1^1$-definable (and the $\Sigma_1^B$-definable)
functions in \VZ(m), and \VACC\
are precisely those in \FACZ(m), and \FACC, respectively.
\end{theorem}

\begin{corollary}
If \VACC\ is finitely axiomatizable, then $\ACC = \ACZ(m)$, for some $m$.
\end{corollary}

\begin{proof}
If \VACC\ is finitely axiomatizable, then by compactness, it is equal to
\[\VZ \cup \{\MOD_i \mid 2 \le i \le m' \},\]
for some $m'$.
Let $m = \lcm\{2, \ldots, m'\}$, then
\[\VZ(m) \vdash \{\MOD_i \mid 2 \le i \le m' \}.\]
Therefore $\VACC = \VZ(m)$, and the conclusion follows from the theorem.
\end{proof}

\begin{corollary}
If $\VACC \vdash \NUMONES$, then $\TCZ = \ACZ(m)$, for some $m$.
\end{corollary}

\subsubsection{Theories for $\NCk$ and $\NC$}

A language is in nonuniform \NCOne\ if it is computable by 
a polynomial-size log-depth family of Boolean circuits.
Here we use uniform \NCOne, which means {\bf Alogtime},
the class of languages computable by alternating Turing machines in log time.
Buss \cite{Buss:87:stoc} shows that the Boolean formula value
problem is complete for {\bf Alogtime}.
In general, for each $k \in \N$ uniform \NCk\ can be considered as the class of relations
which are \ACZ-reducible to the circuit value problem, where the depth of
the circuit is bounded by $(\log{n})^k$.
The function class \FNCk\ consists of functions \ACZ-reducible to
the above problem, or equivalently the functions computable by uniform
polynomial-size $(\log{n})^k$-depth constant-fanins families of Boolean circuits.
Also,
$$\NC = \bigcup_{k \ge 1} \NCk, \qquad \qquad \FNC = \bigcup_{k\ge 1} \FNCk$$

The two-sorted theory \VNCOne\ introduced in \cite{Cook:05:quaderni, Cook:Morioka:04}
is originated from Arai's single-sorted theory \AID\ \cite{Arai:00:apal}.
It is the theory \VZ\ extended by the axiom scheme \SigZBTreeRec,
which essentially exhibits the evaluations of log-depth Boolean circuits
given their specification and inputs.

Informally, consider a log-depth Boolean circuit (i.e., a formula) whose gates can
be numbered such that the input gates are numbered $a, \ldots, 2a-1$,
output gate numbered $1$ and other
internal gates are numbered $2, \ldots, a-1$.
Furthermore, inputs to gate $i$ (where $i < a$) are from gates numbered $2i$ and $2i+1$.
Let the gate $i$ be given by a \SigZB\ formula $\phi(i)[p, q]$ which might have other parameters,
i.e., the intended meaning of $\phi(i)[p, q]$ is the output of
the gate numbered $i$ when its two inputs are $p, q$.
The \SigZBTreeRec\ for $\phi$ explicitly evaluates all the gates of
such circuit when it is given inputs $Z(0), \ldots Z(a)$: for $i < a$, $Z(i)$ is the value
output by gate numbered $i$.
Formally, it is defined as follows:
\begin{equation*}
\exists Z \le 2a \forall i<a[ Z(i+a) \lra \psi(i) \wedge
0 < i \supset (Z(i) \lra \phi(i)[Z(2i), Z(2i+1)]]
\end{equation*}

It has been shown \cite{Cook:05:quaderni, Cook:Morioka:04}
that the $\SigOneOne$-definable functions in \VNCOne\
are precisely the functions in \FNCOne, the function class associated with \NCOne.

It is easy to show that
\VNCOne\ can be axiomatized by \VZ\ and the following single instance of the \SigZBTreeRec\ axioms.
This instance is obtained by replacing $\phi(i)[Z(2i),Z(2i+1)]$ by the 
formula $\Select(W(i), Z(2i), Z(2i+1))$, where $\Select(p,q,r)$ stands for
\begin{equation}\label{e:Select}
(p \wedge (q \wedge r)) \vee (\neg p \wedge (q \vee r))
\end{equation}
Loosely speaking, we think of $W$ as specifying the circuit:
If $W(i)$ holds then the $i$-th gate is a $\wedge$-gate,
otherwise it is a $\vee$-gate.

We will now define the theories characterizing $\NCk$ 
(note that for $k = 1$ we obtain the same theory as \VNCOne,
but we will not prove this fact here).
For each $k$, the complete problem for $\NCk$ is given by a circuit
of depth $O((\log{n})^k)$ and its inputs.
(The function $\log n$ is definable in \IDelZ,
e.g., see \cite{Hajek:Pudlak:93}.)
Consider a circuit of depth $(\log{a})^k$,
where each layer contains at most $(a+1)$ gates.
The layers are indexed according to their depths:
$0$ (input gates), $\ldots$, $(\log{a})^k$ (output gates),
with the outputs of gates on layer $d$ connect to the inputs of
gates on layer $d+1$.
On each layer, the gates are numbered $0, \ldots, a$
(i.e., any gate is indexed by its layer and its position on the layer).

Such circuit can be described by listing the gates together with
their layers index and their inputs gates positions (on the layer below it).
Thus we have a string variable $Y$ which specifies the
wires of the circuit: for $d < \log^k{a}$
and $x, y, z \le a$, $Y^{[d]}(x, y, z)$ holds if and only if
inputs to gate $z$ on layer $d+1$ are from gates $x, y$ on layer $d$.
We also have a string variable $W$ that specifies the type of each gate,
i.e., if $W^{[d]}(z)$ holds then the $z$-th gate on layer $d$ is an $\wedge$-gate, otherwise it is an $\vee$-gate.
The formula $\varphi_{\NCk}(a, X, Y, W, Z)$ below states that
$Z$ evaluates all the gates of the circuit specified by $Y$ and $W$ when
it is given inputs $X(0), \ldots, X(a)$.
In particular, the output of gate $z$ on layer $d$ is $Z^{[d]}(z)$.
The formula $\varphi_{\NCk}(a, X, Y, W, Z)$ is defined to be
\begin{multline*}
\forall d < \log^k{a} \forall z \le a \exists ! x, y \le a Y^{[d]}(x, y, z) \supset \\
 (\forall z \le a Z^{[0]}(z) \lra X(z))\ \wedge \
 \forall d < \log^k{a} \forall x, y, z \le a,\\
Y^{[d]}(x, y, z) \supset (Z^{[d+1]}(z) \lra \Select(W^{[d+1]}(z), Z^{[d]}(x), Z^{[d]}(y)))
\end{multline*}
where $\Select$ is defined in \eqref{e:Select}.

\begin{definition}
Let \ANCk\ denote $\exists Z \varphi_{\NCk}(a, X, Y, W, Z)$.
The theory $\VNC^k$ is \VZ\ extended by the axiom $\ANCk$.
The theory \VNC\ is
$$\bigcup_{k\ge 1} \VNCk$$
\end{definition}

Again, note that in \ANCk, $Z$ can be bounded, therefore \VNCk\ is
equivalent to a theory with bounded axioms.
Proving the first sentence in Theorem~\ref{t:VNCk-boot} below is somewhat
easier than Claim~\ref{t:VTCZ-boot}.

\begin{theorem}
\label{t:VNCk-boot}
For each $k \ge 1$,
$$\VNCk \vdash \forall X \forall Y \forall W\exists Z
\forall w < b \varphi_{\NCk}(a, X^{[w]}, Y^{[w]}, W^{[w]}, Z^{[w]}).$$
A function is in \FNCk\ iff it is $\SigOneOne$-definable in \VNCk.
A function is in \FNC\ iff it is $\SigOneOne$-definable in \VNC.
\end{theorem}

\ignore{
\begin{proof}
Note that the last sentence follows from the second sentence.
Also, by Corollary~\ref{t:VC-define-funcs}, the second sentence
follows from the first.
The proof of the first sentence is straightforward, and we omit the details.
\end{proof}
}

\subsubsection{Theories for \NL, \SL, \L\ and \Ptime}


\NL\ is the class of problems solvable in a nondeterministic Turing machine
in space $O(\log{n})$.
We consider \NL\ as the class of two-sorted relations \ACZ-reducible to
the {\em Graph Accessibility Problem} GAP
(also known as {\em Path}, or {\em Reachability} problem).
This is the problem of deciding whether there is a path from
$s$ to $t$ in a given (directed) graph $G$, where $s, t$ are
the 2 designated vertices of $G$.

We can obtain a theory that characterizes \NL\ by formalizing
the following polytime algorithm that solves GAP.
For each distance $k = 0, 1, \ldots, n-1$ (where $n$ is the number
of vertices in the graph), simply list all vertices that
can be reached from $s$ by paths of length at most $k$.
This enables us to check if $t$ is reachable from $s$ by paths
of length at most $n$,
i.e., if there is a path from $s$ to $t$ in $G$.

Using GAP, the theory \VNL\
is developed in the same style of \VTCZ.
The $\SigOneB$ axiom that formalizes the algorithm solving this problem is called $\LC$
(for Logspace Computation).
In the following definition, $E$ codes a directed graph,
and $Z$ is intended to code the above polytime algorithm.
Here we identify the source $s$ with 0.
Let $\varphi_\LC$ be the following formula
\begin{equation}
\label{e:lc}
\begin{split}
\varphi_\LC(a, E, Z) & \equiv
Z(0,0) \wedge \forall i < a \neg Z(0,i) \wedge \\
& \forall k,i<a,\ Z(k+1, i) \lra [Z(k,i) \vee \exists j<a,\ E(j,i) \wedge Z(k,j)].
\end{split}
\end{equation}
Note that in \eqref{e:lc},
$Z(k,i)$ holds iff there is a path from 0 to $i$ of length at most $k$.

\begin{definition}
[\VNL]
Let $\LC$ denote $\forall a \forall E \exists Z \le (1+\langle a, a \rangle) \varphi_\LC(a, E, Z)$,
Then \VNL\ is the theory \VZ\ extended by the axiom \LC.
\end{definition}

It can be shown directly that the class of \SigOneB-definable functions
in \VNL\ is precisely \FNL, the class of functions whose bit graphs are in \NL.
Here we can show this using Corollary~\ref{t:VC-define-funcs}.
It amounts to showing that $\VNL$ proves the axiom $\AXIOM_{\FGAP}$
(where $\FGAP(a, E)$ is essentially the function whose graph is $\varphi_\LC(e, E, Z)$),
i.e.,
$$\VNL \vdash \forall a \forall b \forall E \exists Z \le \langle b, a, a \rangle
\forall y < b \varphi_{\FGAP}(a, E^{[y]}, Z^{[y]}).
$$
This is analogous to Claim~\ref{t:VTCZ-boot}.
The proof idea is similar; details are omitted.

In the same spirit, a series of theories for
$\L$ (class of problems solvable by a Turing machine in space $O(\log{n})$),
$\SL$ (class of problems solvable by a {\em symmetric nondeterministic} Turing machine in space $O(\log{n})$),
and
$\Ptime$ can be obtained using similar complete problems.
For \L\ the complete problem is GAP restricted to directed graphs whose vertices have
out degree at most 1;
for \SL\ the complete problem is GAP restricted to undirected graphs;
and for \Ptime\ the complete problem is the circuit value problem.

\Remark
It is not a surprise that the theories obtained this way are ``minimal'',
and thus
coincide with a number of existing ``minimal'' theories that characterize
the corresponding classes.
In fact, it can be shown that in case of \L, the theory obtained is actually
Zambella's theory \SRec\ \cite{Zambella:97:apal},
and in case of \Ptime, the theory obtained is the same as \TVZ\ \cite{Cook:05:quaderni}
(and thus the same as \VHorn\ \cite{Cook:Kolokolova:03}).
Thus, these results explicitly exhibit the finite axiomatizability of \SRec\ and \TVZ.
In the case of \NL, it has been shown \cite{Kolokolova:04:thesis} that
\VNL\ is the same as \VKrom\ (see also \cite{Cook:Kolokolova:04}).
In the next section we will show that \VTCZ\ is RSUV isomorphic to
Johannsen and Pollett's ``minimal'' theory \DelCR.

\section{RSUV Isomorphism Between \VTCZ\ And \DelCR}
\label{s:RSUV}

\subsection{The Theory \DelCR}
\label{s:theoryDCR}

The theory \DelCR\ \cite{Johannsen:Pollett:00} is a single-sorted theory
whose \SigOneb\ definable functions are precisely the
(single-sorted) \TCZ\ functions.
It is claimed to be a ``minimal'' theory for \TCZ.
We will show that it is {\em RSUV isomorphic} to our theory \VTCZ.
First, we recall the definition of \DelCR.

The underlying vocabulary of \DelCR\ is
\[\LDelCR = [0, S, +, \cdot, \halfof{x}, |x|, x\# y, \dotminus, \mathit{MSP}; \le]\]
(here $S$ is the successor function,
and $\mathit{MSP}$ stands for most significant bits, $\mathit{MSP}(x,i) = \floor{x/2^i}$).
The theory \DelCR\ is axiomatized by the defining axioms for symbols of \LDelCR, the axiom scheme
\OpenLIND, and the $\DelOneb$ bit-comprehension rule (below).

The defining axioms of the symbols of \LDelCR\ are straightforward.
The axiom scheme \OpenLIND\ can be seen as a scheme of induction
on ``small'' numbers (i.e., $|z|$) for quantifier-free formulas.
Formally, \OpenLIND\ is the set of
\begin{equation}
\label{equation:openlind}
[\varphi(0)\wedge\forall x,\ \varphi(x)\supset\varphi(Sx)]\supset\forall z\varphi(|z|),
\end{equation}
where $\varphi$ is an open formula.
The $\DelOneb$ bit-comprehension rule is defined as follows.%
\label{d:DelOneb-comp-rule}
First, given a
formula $\varphi(i)$ (which might have other free variables),
the comprehension axiom for $\varphi(i)$, denoted by $\COMP{}_{\varphi(i)}(a)$, is
the formula
\[\exists x<2^{|a|} \forall i<|a|\ [\BIT(i,x) \lra \varphi(i)].\]
Here $\BIT(i, x)$ holds if and only if the $i$th bit in the binary
representation of $x$ is 1 (the bits of $x$ are counted from 0 for the lowest order bit).
It is defined by
\begin{align*}
\BIT(i, x) \equiv \mathit{mod2}(\mathit{MSP}(x, i)),
\qquad
\text{ where $\mathit{mod2}(x) = x \dotminus 2\cdot\halfof{x}$.}
\end{align*}
Then, the $\DelOneb$ bit-comprehension rule is the following inference rule:
\begin{align*}
\begin{prooftree}
\varphi(i)\lra\psi(i)
\justifies
\COMP{}_{\varphi(i)}(t)
\end{prooftree}
\end{align*}
where $\varphi$ is a $\SigOneb$ formula, $\psi$ is a $\PiOneb$ formula,
and $t$ is a term.

Note that formally, $\DelCR$ is defined inductively
using the above rule.
More precisely, $\DelCR$ is the smallest theory that contains the
axioms described above, and is closed under the $\DelOneb$ bit-comprehension rule,
i.e., 
if $\varphi(i)\lra\psi(i)$ is in $\DelCR$ for some $\SigOneb$ formula
$\varphi$ and $\PiOneb$ formula $\psi$,
then $\COMP{}_{\varphi(i)}(t)$ is also in $\DelCR$, for any term $t$.
Let $\DelCR_i$ be the sub-theory of \DelCR\ where proving each theorem
of $\DelCR_i$ requires at most $i$ nested applications of the $\DelOneb$
bit-comprehension rule.
Then
\[ \DelCR = \bigcup_{i \ge 0} \DelCR_i.\]
An open question \cite{Johannsen:Pollett:00} is whether $\DelCR = \DelCR_i$
for some constant $i$.
Since \VTCZ\ is finitely axiomatizable,
the RSUV isomorphism between \DelCR\ and \VTCZ\ proved below shows that \DelCR\
is also finitely axiomatizable.
It follows that $\DelCR = \DelCR_i$
for some constant $i$.

Note that there is no side formula in the \DelOneb\ bit comprehension rule,
and thus it is
apparently weaker than the $\DelOneb$ bit-comprehension axiom scheme:
\begin{equation}
\label{e:Del1b-comp}
\forall i(\varphi(i)\lra\psi(i)) \supset \COMP{}_{\varphi(i)}(t)
\end{equation}
where $\varphi$ is a $\SigOneb$ formula and $\psi$ is a $\PiOneb$ formula.
In fact, \cite{Cook:Thapen:04} shows that 
\DelCR\ does not prove the above comprehension axiom scheme
unless RSA can be cracked using probabilistic polynomial time algorithms.

{\bf Remark:} 
We can obtain a single-sorted theory which is equivalent to \DelCR\ as follows.
Let $\calT$ be the theory over the vocabulary $\LDelCR \cup \{\BIT\}$
which is axiomatized by the defining axioms for symbols in $\LDelCR \cup \{\BIT\}$
together with \COMP{\mathbf{\Sigma}_0^b}, i.e., $\COMP{}_{\varphi(i)}(a)$ for
$\mathbf{\Sigma}_0^b$ formulas $\varphi(i)$.
Then the arguments given in Subections 4.2, 4.3 below also show that \calT\ is RSUV isomorphic to \VTCZ.
It follows that \calT\ is a conservative extension of \DelCR.
(A direct proof of this can be obtained by (i) noticing that \BIT\ is definable in \DelCR,
and that $\DelCR$ proves \COMP{\mathbf{\Sigma}_0^b} using the \DelOneb\ bit-comprehension rule;
and (ii) showing that the \DelOneb\ bit-comprehension rule is provable in \calT\
using the same arguments as in Subsection 4.4 below,
and that \OpenLIND\ is provable in \calT\ using \COMP{\mathbf{\Sigma}_0^b} and the axioms for $|x|$ and $\BIT$.)

\subsection{RSUV Isomorphism}

We will prove that \DelCR\ is RSUV isomorphic
\cite{Krajicek:90:apal, Razborov:93:Clote-Krajicek, Takeuti:93:Clote-Krajicek}
to our theory \VTCZ.
A major part of this proof
is in defining multiplication for second sort objects and
proving the commutative and distributive laws.
Here we identify each bounded subset $X$ with the number
\[X(0) 2 ^ 0 + \ldots X(n-1) 2 ^ {n-1},\]
where $n = |X|$.
Then the multiplication function $X\cdot Y$ is defined as the product of
these ``big numbers'' corresponding to $X$ and $Y$.
Note that it is quite straightforward to formalize the conventional 
(polynomial time) algorithm for multiplication in theories that characterize
\Ptime, such as \VOne\ \cite{Cook:02:notes}.
However this is less straightforward in the case of \VTCZ.
Indeed, the conventional algorithm might not be in \TCZ.
Note that the multiplication function is already complete for \TCZ.
Here we define this function in \VTCZ\ using the fact
that computing $Y$ from $X$ in \eqref{e:NUMONES}
(or equivalently the function $\numones$)
is also complete for \TCZ.
In particular, we define multiplication in \VTCZ\ by formalizing a reduction from \numones.
We use the same method as in \cite{Bonet:Pitassi:Raz:00},
where it is shown that the properties of multiplication have small
\TCZ-\Frege\ proofs.

Note that reasoning in \VTCZbar\ is more uniform than in \TCZ-\Frege.
For example, in \VTCZbar\ the ``input bits'' are already ordered
(i.e., the bits in the string $X(n-1) \ldots X(0)$ are numbered),
while this is not the case in \TCZ-\Frege.
Consequently, the definition of the sum of $n$ strings
in \TCZ-\Frege\ does not depend
on the order of the strings (i.e., it is symmetric in terms of the arguments).
In \VTCZ\ proving this independence is a nontrivial task.
It is true, although nontrivial, that each $\SigZB(\LFTCZ)$ theorem of
\VTCZbar\ translates into a family of propositional tautologies having
polynomial-size \TCZ-\Frege\ proofs.
It follows that the proof system
{\em bounded depth} \TCZ-\Frege\ can define string multiplication and prove
its properties using polynomial size proofs.
These bounded depth \TCZ-\Frege\ proofs can be seen as the uniform versions of
the \TCZ-\Frege\ proofs in \cite{Bonet:Pitassi:Raz:00}.

\subsubsection{Outline of the RSUV Isomorphism}
\label{s:RSUVoutline}

We establish the RSUV isomorphism between \DelCR\ and \VTCZ\ by%
\label{d:models-construction}
(a) constructing from each model $\calM$ of \DelCR\ a model $\calN$ of
\VTCZ\ whose second sort universe is the universe $M$ of $\calM$,
and whose first sort universe is the subset $\log(M) = \{|u| \mid u \in M\}$;
and
(b) constructing from each model $\calN$ of \VTCZ\ a model $\calM$ of
\DelCR\ whose universe is the second sort universe of $\calN$.
These constructions have the property that if we follow (a) to
get a model $\calN$ of \VTCZ\ from the model $\calM$ of \DelCR,
and then follow (b) to get a model $\calM'$ of \DelCR\ from $\calN$,
then $\calM'$ and $\calM$ are isomorphic.
Similarly, 
if we start with a model $\calN$ of \VTCZ\ and follow (b) to get
a model $\calM$ of \DelCR, then follow (a) to get $\calN'$ from $\calM$,
then $\calN$ and $\calN'$ are isomorphic.

Corresponding to (a) there is a syntactic translation sending a
sentence $\varphi$ in the language of \VTCZ\ to an equivalent sentence
$\varphi^\flat$ in the language of \DelCR\, and corresponding to
(b) there is a syntactic translation sending a sentence $\psi$
of \DelCR\ to an equivalent sentence $\psi^\sharp$ in the language of \VTCZbar.
(The $\flat$ and $\sharp$ notation is from \cite{Razborov:93:Clote-Krajicek}.)
Recall from Corollary \ref{t:VTCZbar-VTCZ} that \VTCZbar\
is a conservative extension of \VTCZ.  These translations can
be pictured as follows:
\begin{center}
\begin{tabular}{ccc}
\DelCR  &  $\simeq$ & \VTCZ \\
$\calM$ & $\rightharpoonup$ & $\calN$ \\
$\varphi^\flat$ & $\leftharpoonup$ & $\varphi$ \\ \hline
$\calM'$ & $\leftharpoonup$ & $\calN'$  \\
$\psi$  & $\rightharpoonup$ & $\psi^\sharp$ 
\end{tabular}
\end{center}
The construction in (a) is straightforward.
Let $\calM$ be a model of \DelCR\ with universe $M$,
we construct a model $\calN$ of $\VTCZ$ as follows.
To get the second sort universe of $\calN$,
we simply identify each number $a \in M$ with the subset $X$
(bounded by $|a|$) of those 
$i$ such that the $i$th bit in the binary representation of $a$ is 1.
The symbols of \VTCZ\ are interpreted accordingly, i.e,
$0, 1, +, \cdot, =_{1}$ and $\le$ are interpreted as in $\calM$
(restricted to $\log(M)$), and
\begin{align*}
|X| = |a|,\ 
\text{ and }
i \in X \Lra \BIT(i, a) = 1
\qquad
\text{ if $X = \{i \mid \BIT(i, a) = 1\}$},\\
X = _2 Y \text{ if they are mapped from the same number $a \in M$.}
\end{align*}
It remains to show that
the axioms of \VTCZ\ hold in $\calN$.
We use the fact that each $\SigZB$ formula $\varphi$ of \VTCZ\
translates to the formula $\varphi^\flat$ which is provably equivalent
in \DelCR\ to a $\mathbf{\Sigma}_0^b$ formula.
Since $\calM$ is a model of \DelCR, it is easy to check that the
axioms in \BASIC{2} are satisfied in $\calN$.
The axiom scheme $\COMP{\SigZB}$ are satisfied in $\calN$
since $\calM$ satisfies the \DelOneb\ bit-comprehension rule.
Now we show that \NUMONES\ holds in $\calN$.
Let $a \in M$, and $a_{n-1}\ldots a_0$ be the binary representation of $a$.
We need to get the ``counting array'' for the set 
\[X = \{i \mid \BIT(i, a) = 1\}.\]
Let $a' \in M$ whose binary representation is $a_{n-1}0\ldots 0 a_{n-2}0\ldots 0a_0$,
where every block of 0's has length $(1+|n|)$.
Let $b \in M$ with the binary representation $10\ldots 0 10\ldots 01$ ($n$ 1's, and each block of 0's
has length $1+|n|$).
Note that $a'$ and $b$ exist in $M$ by the \DelOneb\ bit-comprehension rule.
It is straightforward that
the counting array for $X$ can be extracted from the product $a'\cdot b$.

The construction in (b) is done by reversing the above construction.
Suppose that $\calN$ is a model of $\VTCZ$.
We can view $\calN$ as a model of $\VTCZbar$, where the symbols
of $\LFTCZ$ are interpreted according to their defining axioms given in Definition~\ref{d:LFTCZ}.
We construct a model $\calM$ for \DelCR\ by interpreting
each second sort object $X$ of $\calN$ as the number
\begin{equation}
\label{e:set-num}
X(0) 2^0 + \ldots + X(n-1) 2^{n-1},
\end{equation}
(i.e., the number whose binary representation is $X(n-1) \ldots X(0)$)
where $n = |X|$.
All symbols of \LDelCR\ except for $\cdot$ are interpreted in a straightforward manner.
For example, if $a$ is the number in $\calM$ with the value
from \eqref{e:set-num}, for some second sort object $X \in \calN$,
then $|a| = |X|$ (more precisely, there is a second sort object
$Z \in \calN$ such that in $\calN$, $|X| = Z(0) 2^0 + \ldots + Z(m-1) 2^{m-1}$, where $m = |Z|$,
and $|a|$ is the number associated with $Z$).
The axioms of \DelCR\ describing these symbols hold in $\calM$
because their translations (except those involving $\cdot$)
are easy theorems of $\VTCZbar$.
(We will present a proof of the associativity of string addition in Appendix~\ref{s:Interp-Addition}.)
It remains to (i) interpret $\cdot$, and prove its properties in $\calM$,
and
(ii) show that other axioms of $\DelCR$ are satisfied in $\calM$.
For (ii), the axiom scheme \OpenLIND\ holds in $\calM$,
since $\mathbf{Open}$-$\mathbf{IND}$ holds in $\VTCZbar$.
Therefore we will present only the proof that the
\DelOneb\ bit-comprehension rule is satisfied in $\calM$.

\subsection{Interpreting Multiplication for the Second Sort Objects in \VTCZbar}

Now we need to define the ``string multiplication'' function $X \cdot_2 Y$
(we will simply write $X \cdot Y$),
which is the binary representation of the product of the two numbers
corresponding to $X$ and $Y$ by the mapping \eqref{e:set-num}.
It is known that the complexity of computing $X\cdot Y$ is \ACZ\ complete
for \TCZ\ \cite{Chandra:Stockmeyer:Vishkin:84}.
Thus our task is to formalize in \VTCZ\ a \TCZ\ algorithm computing this product.
This can be reduced to computing the sum of $n$ strings.
The ``school algorithm'' is to write down the strings
and sum up the bits in the same columns, starting from the lowest
order bits, with carries from the previous columns.
However, it might not be possible to formalize this algorithm in \VTCZbar,
and we will formalize the algorithm from \cite{Bonet:Pitassi:Raz:00},
where it is shown that multiplication can be defined in \TCZ-\Frege.

\subsubsection{Adding $n$ Strings}

Suppose that we are to add $n$ strings, each of length $\le m$
(written as a table of $n$ rows and $m$ columns).
The methods from \cite{Bonet:Pitassi:Raz:00} is to divide the $m$ columns
into $2k$ blocks, each consisting of $\ell$ columns. 
(Thus each block has $n$ substrings of length $\ell$.)
The numbers $k$ and $\ell$ are chosen so that
the sums of the substrings in the $2k$ blocks can be computed concurrently.
It remains to use these sub-sums to obtain the desired sum;
a further requirement for $k$ and $\ell$ is that this last step can
be carried out efficiently.

More precisely let $\ell = 1+\ceil{\log n}$ and $k = \ceil{m/2\ell}$.
Notice that the sum of the $n$ substrings in each block is a string of length bounded by
$2\ell$, or equivalently a number (i.e., first sort object of $\calN$)
which is $\le 2^{2\ell} \le 4n^2$.
Let $b_0, \ldots, b_{2k-1}$ be the sub-sums.
The sum of the original $n$ strings is computed from
$2k$ such ``short'' strings by
first ``concatenating'' $b_0, b_2, \ldots, b_{2k-2}$
and ``concatenating'' $b_1, b_3, \ldots, b_{2k-1}$
(i.e., concatenating the binary string representations of $b_0, b_2, \ldots, b_{2k-2}$,
and concatenating the binary string representations of $b_1, b_3, \ldots, b_{2k-1}$),
then adding the 2 resulting strings together.

Formally,
suppose that the $n$ strings are represented
as $n$ rows $Z^{[0]}, \ldots, Z^{[n-1]}$ in an array $Z$ (using the pairing function).
Our goal is to compute their sum $\Sum(n,m,Z)$
as a string function of $n, m, Z$.%
\footnote{Here $n, m$ indicate the ``size'' of $Z$, i.e., it has $n$ rows, each of length $\le m$.}
Note that if
for each $i$, $0\le i<m$, $c_i$ is the total number of bits in
the $i$th column of $Z$, then
\[\Sum(n,m,Z) = \sum_{i=0}^{m-1} 2^i c_i.\]

We will ``store'' $c_i$'s in a string $W$, and then
define $\Sum(n,m,Z)$ as a function\\ $\Sum'(m, n, W)$.%
\footnote{Note the difference in ordering of $n$ and $m$ as arguments in $\Sum$ and $\Sum'$.}
Here $W$ is a \TCZ\ string function of $n, m$ and $Z$:
it has $m$ rows, and the row $W^{[i]}$ of $W$ has length exactly $c_i$.
It can be defined as follows.
First, let $\Zb$ be the transpose of $Z$:
for $0 \le i < m$,
\[|\Zb^{[i]}| \le n \wedge \forall j < n\ \Zb^{[i]}(j) \lra Z(j, i).\]
Then the total number of bits in the $i$th column of $Z$ is
exactly $\numones(n,\Zb^{[i]})$, the total number of bits in the $i$th row of $\Zb$.
Now $W = \AddCols(n, m, Z)$ where $\AddCols(n, m, Z)$ is defined by
\begin{equation}
\label{e:AddCols}
|\AddCols(n, m, Z)| \le \langle m, n \rangle
\wedge
\forall i < m, j < n\ \AddCols(n, m, Z)(i, j) \lra
j < \numones(n,\Zb^{[i]}).
\end{equation}

We need to compute
\begin{equation}
\label{e:Sum'}
\Sum'(m, n, W) = \sum_{i=0}^{m-1} 2^i |W^{[i]}|,
\end{equation}
where $n$ is a bound for $|W^{[0]}|, \ldots, |W^{[m-1]}|$:
$|W^{[i]}| \le n$ for $0 \le i < m$.

Notice that the number functions $2^x$
where $x < |a|$ for some number $a \in \calN$,
and $\log{x}$, are in \FTCZ\ (in fact, they are in \FACZ
\cite{buss98bookB,Cook:02:notes}).
Let $\ell$ and $k$ be as in the above discussion, i.e.,
\[\ell = 1+\ceil{\log n}, \qquad k = \ceil{m/2\ell}.\]

Write $c_i$ for $|W^{[i]}|$, for $0 \le i < m$.
Divide $c_{m-1}, \ldots, c_0$ into $2k$ blocks of length $\ell$ each:
\[c_{2k\ell - 1}, \ldots, c_{(2k-1)\ell};
\qquad
\ldots;
\qquad
c_{2\ell-1}, \ldots, c_\ell;
\qquad
c_{\ell-1}, \ldots, c_0.\]
For $0\le i<2k$, we will define $b_i$ to be the sum of the $i$th block, $b_i = \sum_{j=0}^{\ell-1} 2^j c_{i\ell+j}$.
Formally, this is a number function of $W, i$ and $\ell$,
i.e.,
$b_i = \ssum(W, i\ell, \ell)$ where
\[\ssum(W, a, \ell) = \sum_{j=0}^{\ell-1} 2^j c_{a+j}\]
(the sum of the block of length $\ell$, starting from $a$).
Here, $\ssum$ can be defined using $\numones$:
$\ssum(W, a, \ell)$ is the number of bits in the ``long''
string $Y$,
\[\ssum(W, a, \ell) = \numones(|Y|, Y),\]
where $Y$ consists of 
$1$ substring of $c_a$ 1's; $2^1$ substrings, each of $c_{a+1}$ 1's; $\ldots$;
$2^{\ell-1}$ substrings, each of $c_{a+\ell-1}$ 1's.
Obviously, we can define such $Y$ as an \ACZ\ function of $W$:
\[|Y| \le 2^\ell n \wedge \forall j<\ell\forall u<2^j \forall v<n\ [Y((2^j-1)n + un + v) \lra v<c_{a+j}],\]
(note that $n$ is a bound for $c_a, \ldots, c_{a + \ell-1}$).
Also, \VTCZbar\ proves the following properties of $\ssum$:
\[\ssum(W,a,0) = c_a,
\qquad
\ssum(W,a,\ell+1) = \ssum(W,a,\ell) + 2^\ell c_{a+\ell},
\qquad
\ssum(W,a,\ell) < n 2^\ell.\]
In particular, we have $b_i < 2^{2\ell}$, for $i < 2k$.

Let 
\[L = \sum_{i=0}^{k-1}2^{2i\ell}b_{2i},
\qquad
H = \sum_{i=0}^{k-1}2^{(2i+1)\ell}b_{2i+1}.\]
Since $b_i < 2^{2\ell}$ for $i < 2k$, $L$ and $H$
can be computed simply by concatenating the binary representations
of $b_0, b_2, \ldots, b_{2k-2}$ and
$b_1, b_3, \ldots, b_{2k-1}$, respectively.
(More precisely, we may have to pad each $b_i$ with leading 0's
to make them of length exactly $2\ell$,
and then concatenate these strings of equal length.)

Now
\[\Sum'(m, n, W) = \sum_{i = 0}^{2k-1} 2^{i\ell} b_i = L + H.\]
As a result, $\Sum(n, m, Z) = \Sum'(m, n, W)$ is a function of \LFTCZ.
Thus we can define $X \cdot Y$ as follows.

Given $X$ and $Y$, let $X \otimes Y$ be the ``table'' that we use
in the ``school algorithm'' to multiply $X$ and $Y$,
$X \otimes Y = Z$ where
\begin{equation}
\label{e:otimes}
|Z| \le (|X| + |Y|)|Y| \wedge \forall x < |X| \forall y < |Y|,\
Z(y, x + y) \lra [X(x) \wedge Y(y)].
\end{equation}
($Z$ has $|Y|$ rows, each is of length $\le |X| + |Y|$.)
Then $X \cdot Y = \Sum(|Y|, |X| + |Y|, Z)$.
It follows that $X \cdot Y$ is a function of $\LFTCZ$.
It remains to prove the properties of this function, i.e.,
it is commutative, and distributive over $X + Y$.

\subsubsection{Proving Properties Of $X \cdot Y$}

First we need to show that $\cdot_2$ is commutative.

\begin{lemma}
\label{t:commutative}
$\VTCZbar \vdash X \cdot Y = Y \cdot X$.
\end{lemma}

\begin{proof}
Recall that we define $\Sum(n, m, Z) = \Sum'(m, n, W)$,
where $W = \AddCols(n, m, Z)$ which is defined in \eqref{e:AddCols}.
Thus it suffices to show that
\begin{equation}
\label{e:otimes-commutative}
\AddCols(|Y|, m, X \otimes Y) = \AddCols(|X|, m, Y \otimes X),
\end{equation}
where $m = |X| + |Y|$.

Let $Z_1 = X \otimes Y$ and $Z_2 = Y \otimes X$.
Notice that for $i < m$ the column $\Zb_1^{[i]}$ of $Z_1$
and column $\Zb_2^{[i]}$ of $Z_2$ are just permutation
of each other.
In particular,
$|\Zb_1^{[i]}|, |\Zb_2^{[i]}| \le i + 1$, and
\[\Zb_1^{[i]}(y) \lra y \le i \wedge Y(y) \wedge X(i - y),
\qquad
\Zb_2^{[i]}(x) \lra x \le i \wedge X(x) \wedge Y(i - x),\]
and hence $\Zb_1^{[i]}(y) \lra \Zb_2^{[i]}(i - y)$, for $y \le i$. 
To prove \eqref{e:otimes-commutative}, we will show that
for $i < m$,
$\Zb_1^{[i]}$ and $\Zb_2^{[i]}$ have the same number of elements, i.e.,
\[\numones(i,\Zb_1^{[i]}) = \numones(i,\Zb_2^{[i]}).\]
It suffices to prove more generally
that if there is an one-one mapping between 
$\Zb_1{[i]}$ and $\Zb_2{[i]}$, then they have the same number of elements.
This is proved in the next lemma.
\end{proof}

In the following lemma,
suppose that there is an one-one mapping (specified by $M$) between the initial segments
$\{ i \mid i \in X \wedge i < \ell \}$ and $\{ j \mid j \in Y \wedge j < \ell\}$
of $X$ and $Y$ respectively.
Then these initial segments have the same number of elements.

\begin{lemma}
\label{t:permutation}
Let $\ell, X, Y, M$ be such that
\begin{equation}
\label{e:permutation}
\forall i<\ell\exists !j<\ell M(i,j)\ \wedge \ \forall j<\ell\exists !i<\ell M(i,j),
\ \ 
\text{and}
\ \ 
\forall i<\ell,\ X(i)\lra \exists j<\ell (M(i,j)\wedge Y(j)).
\end{equation}
Then, $\VTCZ\vdash \numones(\ell,X) = \numones(\ell,Y)$.
\end{lemma}

\begin{proof}
First, from \eqref{e:permutation} it is easy to see that
\[\forall j < \ell,\ Y(j) \lra \exists i < \ell (X(i) \wedge M(i, j)).\]
Let $Z$ be the string such that
$Z^{[k]}$ is the image of the initial segment $\{ i \mid i \in X \wedge i < k\}$
of $X$, i.e.,
$$\forall k<\ell\forall j<\ell\ [Z^{[k]}(j) \lra \exists i< k (M(i,j)\wedge X(i))].$$
Then, $Z^{[\ell]} = Y$.
We can prove by induction on $k$ that $\numones(k,X) = \numones(\ell,Z^{[k]})$.
Consequently, $\numones(\ell,X) = \numones(\ell,Y)$.
\end{proof}

Now we will show that $\cdot_2$ is distributive over $+_2$.

\begin{lemma}
\label{t:distributive}
$\VTCZbar\vdash X\cdot(Y+Z) = X\cdot Y + X\cdot Z.$
\end{lemma}

\begin{proof}
It suffices to prove
\begin{equation}
\label{e:distri-hyp}
  X^{<i} \cdot (Y+Z) = X^{<i} \cdot Y + X^{<i} \cdot Z
\end{equation}
by induction on $i$, where $X^{<i}$ is the $i$ low-order bits of $X$;
that is
\[
   X^{<i} = \{ j<i \mid X(j)  \}.
\]
The base case follows from the fact that $\setZ \cdot Y = \setZ$, which
can be proved from the definition of $X\cdot Y$.

For the induction step there are two cases:  $X^{<i+1} = X^{<i}$
and $X^{i+1}= X^{<i}+ \{i\}$.  Since the first case is trivial,
we consider the second case.  To simplify notation, we will write
$X$ for $X^{<i}$ and write $X'$ for $X+\{i\} = X^{<i+1}$.

Thus our task is to prove in $\VTCZ$
\begin{equation}
\label{e:distr-indstep}
    X'\cdot (Y+Z) = X'\cdot Y + X'\cdot Z
\end{equation}
from the induction hypothesis
\[
  |X|\le i \ \wedge \  X\cdot (Y+Z) = X\cdot Y + X\cdot Z.
\]
We need the following fact, which we prove
below:
\begin{equation}
\label{e:distri-2}
\VTCZbar \vdash |X| \le i \supset
          (X + \{i\}) \cdot Y = X \cdot Y + \{i\} \cdot Y.
\end{equation}
From the definition of $X\cdot Y$ we have
\begin{equation}
\label{e:distri-1}
X \cdot \{i\} = \{x + i \mid x \in X\}
\end{equation}
From the commutativity of $\cdot_2$ (Lemma~\ref{t:commutative}),
the associativity of $+_2$ (Lemma~\ref{t:associative}),
and \eqref{e:distri-1} we can derive that
\begin{equation}
\label{e:distri-1a}
\{i\} \cdot (Y + Z) = (Y + Z) \cdot \{i\} = Y \cdot \{i\} + Z \cdot \{i\} 
= \{i\} \cdot Y + \{i \} \cdot Z.
\end{equation}
Now we prove (\ref{e:distr-indstep}) as follows, using associativity
and commutativity of $+_2$.
\begin{eqnarray*}
X'\cdot (Y+Z) & = & X\cdot(Y+Z) + \{i\}\cdot(Y+Z) \qquad
                                  \mbox{[by (\ref{e:distri-2})]}  \\
& = &  X\cdot Y + X\cdot Z + \{i\}\cdot(Y+Z)  \qquad
                                  \mbox{[by Ind Hyp]}  \\
& = &  X\cdot Y + X\cdot Z + \{i\}\cdot Y + \{i\}\cdot Z  \qquad
                           \mbox{[by (\ref{e:distri-1a})]}  \\
& = &   X'\cdot Y + X'\cdot Z \qquad   \mbox{[by (\ref{e:distri-2})].}  
\end{eqnarray*}

It remains to prove \eqref{e:distri-2}.
Using the commutativity of $\cdot_2$,
rewrite the equality in \eqref{e:distri-2} as
\begin{equation}
\notag
Y \cdot (X + \{i\}) = Y \cdot X + Y \cdot \{i\}.
\end{equation}
To prove this, it suffices to prove that
\[\Sum(i, i + |Y|, Y \otimes (X + \{i\})) = \Sum(|X|, |X| + |Y|, Y \otimes X) + Y \cdot \{i\}.\]
Notice that since $|X| \le i$, the ``table'' $Z_1 = Y \otimes (X + \{i\})$
is exactly $Y \otimes X$ appended with an additional row
\[Z_1^{[i]} = \{y + i \mid y \in Y\} = Y \cdot \{i\}.\]
Therefore \eqref{e:distri-2} follows from the next lemma.
\end{proof}

\begin{lemma}
Suppose that $|Z^{[i]}| \le m$, for $0 \le i \le n$.
Then $\VTCZbar \vdash \Sum(n+1, m, Z) = \Sum(n, m, Z) + Z^{[n]}$.
\end{lemma}

\begin{proof}
We have defined $\Sum$ using $\Sum'$.
(Recall the definition of $\Sum'$ in \eqref{e:Sum'}.)
We need to show that
\[\Sum'(m, n + 1, W) = \Sum'(m, n, W_1) + Z^{[n]}\]
where 
\[W_1 = \AddCols(n, m, Z),
\qquad
W = \AddCols(n + 1, m, Z),\]
It is straightforward that for $i < m$,
\begin{equation}
\label{e:Sum'-prop-1}
|W^{[i]}| =
\begin{cases}
|W_1^{[i]}| + 1 & \text{ if } i \in Z^{[n]}\\
|W_1^{[i]}| & \text{ if } i \not\in Z^{[n]}.
\end{cases}
\end{equation}
We will prove by induction on $m' \le m$ that
\begin{equation}
\label{e:Sum'-prop-2}
\Sum'(m', n + 1, W) = \Sum'(m', n, W_1) + (Z^{[n]})^{< m'},
\end{equation}
where $X^{< z}$ is the initial segment $\{x \mid x \in X \wedge x < z\}$
of $X$.

For the base case, \eqref{e:Sum'-prop-2} obviously holds when $m' = 0$.

For the induction step, suppose that \eqref{e:Sum'-prop-2} holds for some $m' < m$.
Note that by the definition of $\Sum'$ \eqref{e:Sum'}, intuitively,
\[\Sum'(m' + 1, n + 1, W) = \Sum'(m', n + 1, W) + 2^{m'} |W^{[m']}|.\]
Formally, let $\ToString(c, \ell)$ be the set 
\[\ToString(c, \ell) = \{i + \ell \mid \BIT(i, c)\}.\]
Then we can show from the definition of $\Sum'$ that
\[\Sum'(m' + 1, n + 1, W) = \Sum'(m', n + 1, W) + \ToString(c_{m'}, m'),\]
(where $c_{m'}$ stands for $|W^{[m']}|$).
Similarly,
\[\Sum'(m' + 1, n, W_1) = \Sum'(m', n, W_1) + \ToString(c^1_{m'}, m'),\]
where $c^1_{m'}$ stands for $|W_1^{[m']}|$.

By the induction hypothesis, it remains to show that
\[\ToString(c_{m'}, m') + (Z^{[n]})^{< m'} = \ToString(c^1_{m'}, m') + (Z^{[n]})^{< m' + 1}.\]
This follows from \eqref{e:Sum'-prop-1} and the definition of $\ToString$.
\end{proof}

\subsection{Interpreting the $\DelOneb$ Comprehension Rule in \VTCZ}

Recall the definition of the $\DelOneb$ bit-comprehension rule
(in single-sorted logic) in section \ref{s:theoryDCR}.
Note that this rule specifies an inductive definition of $\DelCR$.
In order to show that $\calM$ is a model of \DelCR,
where $\calM$ is the structure
constructed from a model $\calN$
of \VTCZ\ as discussed in (b) of section \ref{s:RSUVoutline},
we will show that \VTCZ\ (and \VTCZbar)
satisfies the $\gDelOneB$ {\em comprehension rule},
the two-sorted version of the
$\DelOneb$ bit-comprehension rule.
This rule for two-sorted theories is obtained from the
$\DelOneb$ bit-comprehension
rule using the syntactic translation $\psi \rightharpoonup \psi^\sharp$
described in section \ref{s:RSUVoutline}.

Recall the definition of $\gSigOneB$ and $\gPiOneB$ formulas given
in section \ref{s:Syntax-Semantics}.  Note that
under the translation $\psi \rightharpoonup \psi^\sharp$,
$\SigOneb$ and $\PiOneb$
formulas translate to formulas equivalent to those in
$\gSigOneB$ and $\gPiOneB$, respectively.

The $\gDelOneB$ comprehension rule is defined as follows.

\begin{definition}
A (two-sorted) theory $\calT$ over $\calL$
is said to admit the $\gDelOneB(\calL)$ comprehension rule if
whenever
\begin{equation}
\notag
\calT \vdash \forall z<b,\ \varphi(z) \lra \psi(z),
\qquad
\text{then}
\qquad
\calT\vdash \exists X\le b\forall z<b,\ X(z)\lra \varphi(z)
\end{equation}
where $\varphi$ is a $\gSigOneB(\calL)$ formula, and $\psi$ is a $\gPiOneB(\calL)$ formula
($\varphi$ and $\psi$ may have other free variables).
\end{definition}

We omit $\calL$ from $\gDelOneB(\calL)$ when it is clear from context.

This rule is apparently weaker than the $\gDelOneB$ comprehension axiom.
In particular, \VTCZ\ may not prove the axiom. 

Our task for this
section to prove the following theorem.

\begin{theorem}
\label{t:DelOnecomp}
\VTCZ\ and \VTCZbar\ admit the $\gDelOneB$ comprehension rule.
\end{theorem}

Note that it suffices to prove the theorem for \VTCZbar, since
this theory is conservative over \VTCZ.
We will use the $\gSigOneB$ replacement rule, defined as follows.

\begin{definition}
\label{d:replacement-rule}
A (two-sorted) theory $\calT$ over $\calL$
is said to admit the $\gSigOneB(\calL)$ replacement rule if
whenever
\begin{equation}
\notag
\calT\vdash\forall z<b\exists Z<b\varphi(z,Z),
\qquad
\text{then}
\qquad
\calT\vdash\exists W<\langle b,b\rangle\forall z<b\ \varphi(z,W^{[z]}),
\end{equation}
for any $\gSigOneB(\calL)$ formula $\varphi$ which may contain other free variables.
\end{definition}

Note that if $\calT$ admits the $\gSigOneB$ replacement rule, then
each $\gSigOneB$ theorem of $\calT$ is provably
equivalent in $\calT$ to a $\SigOneB$ formula.

\begin{lemma}
\label{t:repl-comp-rule}
If the theory $\calT$ (extending $\VZ(\Row)$) proves $\COMP{\SigZB(\calL)}$
and admits the $\gSigOneB(\calL)$ replacement rule,
then it also admits the $\gDelOneB(\calL)$ comprehension rule.
\end{lemma}

\begin{proof}
Suppose that
\[\calT \vdash \forall z<b,\ \varphi(z) \lra \psi(z),\]
where $\varphi$ is a $\gSigOneB(\calL)$ formula,
and $\psi$ is a $g\PiOneB(\calL)$ formula
which may have other free variables.
Then
\[\calT \vdash \forall z < b,\ \varphi(z) \vee \neg \psi(z).\]
Therefore
\[\calT \vdash \forall z < b \exists Z \le 1,\
[Z(0) \wedge \varphi(z)] \vee [\neg Z(0) \wedge \neg \psi(z)].\]
Now $\theta(z, Z) \equiv [Z(0) \wedge \varphi(z)] \vee [\neg Z(0) \wedge \neg \psi(z)]$
is equivalent to a $\gSigOneB$ formula.
Since $\calT$ admits the $\gSigOneB(\calL)$ replacement rule,
\[\calT \vdash \exists W<\langle b,b\rangle \forall z<b \
\theta(z, W^{[z]}).\]

Let $X$ be defined by $\COMP{\SigZB(\calL)}$:
\[|X| \le b \wedge \forall z < b,\ X(z) \lra W^{[z]}(0).\]
Then obviously $\forall z < b,\ X(z) \lra \varphi(z)$.
\end{proof}

Now Theorem \ref{t:DelOnecomp} follows from the following lemma.

\begin{lemma}
\label{t:VTCZbar-repl-rule}
\VTCZbar\ admits the $\gSigOneB(\LFTCZ)$ replacement rule.
\end{lemma}

\begin{proof}
Suppose that 
\[\VTCZbar \vdash \forall z<b\exists Z<b\varphi(z,Z),\]
for some $\gSigOneB(\LFTCZ)$ formula $\varphi(z, Z)$.
By the $\gSigOneB$ Witnessing Theorem (Theorem~\ref{t:witnessing})
there is a function $F(z)$ of $\LFTCZ$ such that
\[\VTCZbar \vdash \forall z<b \varphi(z, F(z)).\]
Let $G$ be defined as follows:
\[|G| \le \langle b, b \rangle \wedge \forall z < b G^{[z]} = F(z).\]
Then we have
\[\forall z < b \varphi(z, G^{[z]}).\]
Also, $G$ is a function of $\LFTCZ$.
Therefore $\VTCZbar \vdash \exists W \le \langle b, b \rangle \forall z < b \varphi(z, W^{[z]})$.
\end{proof}

\section{Conclusion}
\label{s:conclusion}

We show (Theorem~\ref{t:TC0-threshold-quant})
that the \TCZ\ relations are precisely those represented by \SigZBTh\ formulas.
We also present the finitely axiomatizable, second-order theory \VTCZ\ which
characterizes \TCZ\ in the same way that Buss's theory $\mathbf{S}^1_2$
characterizes polynomial time.
Our characterization of \TCZ\ by the theory \VTCZ\ is based on the fact that counting
the number of 1 bits in a string is complete for \TCZ\
rather than the ``hidden power'' of the multiplication
function (which is also complete for \TCZ) usually present
{\em a priori} in first-order theories.

We show that a number of combinatorial problems are provable in \VTCZ.
In particular, we show that \VTCZ\ is RSUV isomorphic to Johannsen and Pollett's
``minimal'' theory \DelCR.
The main part of proving this RSUV isomorphism is in defining (string)
multiplication and proving its properties.
The RSUV isomorphism between \VTCZ\ and \DelCR\ shows that $\DelCR = \DelCR_i$ for some constant $i$,
answering a question in \cite{Johannsen:Pollett:00}.

In addition, we show that a form of the Pigeonhole Principle is provable in \VTCZ.
In \cite{Buss:03:submitted:hex} Buss shows that the STCONN tautologies
(and thus the HEX tautologies) have polynomial size constant depth \TCZ-\Frege\ proofs.
It can be seen that his arguments can be formalized in our theory \VTCZ.
The proofs of these principles in \VTCZ\ are more uniform that the \TCZ-\Frege\ proofs.

In \cite{Hesse:01:icalp}, it is shown that division is in uniform \TCZ.
Thus the (string) division function might be $\SigOneB$-definable in \VTCZ.
An interesting problem is to formalize the algorithm of
\cite{Hesse:01:icalp} in \VTCZ.
\footnote{This problem was suggested by Albert Atserias.}

We are also able to generalize the method used in developing \VTCZ\
to obtain a scheme of theories characterizing a number of other subclasses of \Ptime.
Our work follows the program outlined in \cite{Cook:02:notes,Cook:05:quaderni},
which proposes defining and studying second-order
theories and propositional proof systems associated with
various complexity subclasses of \Ptime.
We have not treated the connection with propositional proof systems here,
but this is the subject of ongoing investigation.
By translating proofs in our theories to the
{\em quantified propositional proof system} \G\
\cite{Krajicek:Pudlak:90, Morioka:05:thesis}, our theories should
correspond to fragments of \G\
which lie between {\em bounded depth} \Frege\ and $\G_1^*$.
In this line, the fragment for \NCOne\ which is different from 
\Frege\ is called \GZ\ in 
\cite{Cook:Morioka:04, Morioka:05:thesis}.
A proof system associated with \L\ can be found in
\cite{Perron:05:thesis}.
Investigation into proof systems corresponding to other classes
is ongoing.

The other direction in the tight connection between first-order theories
and the propositional proof systems is the Reflection Principle:
Each theory proves the soundness of the corresponding proof system.
For example, \PV\ proves the soundness of {\em extended Resolution},
and in fact extended Resolution is the strongest proof system whose
soundness is provable in \PV\ \cite{Cook:75:stoc}.
In general, the \SigZB\ consequences of the theory corresponding to
a complexity class
can be axiomatized by formalizing the soundness
of the corresponding proof system.
Because the bounded depth \Frege\ systems form a proper hierarchy
\cite{Krajicek:94},
this seems to imply that the \SigZB\ consequences of \VZ\ are
not finitely axiomatizable.
Similar (but conditional) results for \VTCZ\ should also hold.

Another interesting issue is to compare various theories
that characterize the same class.
For example, it is possible that $\VNC \subsetneq \UOneOne$,
where \UOneOne\ \cite{Cook:05:quaderni} is a theory that also characterizes \NC.

\section*{Acknowledgment}

We would like to thank the referees for very helpful comments.
We also thank Christ Pollett for clarifying the proofs in \cite{Johannsen:Pollett:98},
and Alan Skelley for helpful comments.

\bibliographystyle{alpha}
\bibliography{ntp}

\appendix

\section{Interpreting Addition for the Second Sort Objects in \VZbar}
\label{s:Interp-Addition}

We define the ``string addition'' function $X +_2 Y$
(we will simply write $X + Y$, the meaning will be clear from the context),
which is the binary representation of the sum of the two numbers
corresponding to $X$ and $Y$ by the mapping \eqref{e:set-num}.

We will show that the string function $X + Y$ can be defined
in any model of $\VZ$.
Here addition is defined using the conventional algorithm,
i.e., adding digits of the same order in $X$ and $Y$ together
with carries from the previous result.
More precisely, let
$\varphi_+(i,X,Y)$ represent the carry at the
bit position $i$ when adding $X$ and $Y$.
Then the $i$th bit of the sum $X+Y$ is
\[(X+Y)(i) \Lra X(i)\oplus Y(i)\oplus\varphi_+(i,X,Y)\]
(Here $\oplus$ stands for {\it exclusive or},
i.e.,
$p \oplus q \lra (p \wedge \neg q) \vee (\neg p \wedge q)$.)

\begin{definition}
Let $\varphi_+(i,X,Y)$ be the $\SigZB$ formula
\begin{equation}
\label{e:carry}
\exists j<i,\ X(j)\wedge Y(j)\wedge
\forall \ell<i[j< \ell\supset X(\ell)\oplus Y(\ell)].
\end{equation}
Then $X + Y$ is defined as follows:
\[|X + Y| \le |X| + |Y|\
\wedge\
\forall i < |X| + |Y|,\ (X+Y)(i)\lra[X(i)\oplus Y(i)\oplus\varphi_+(i,X,Y)].\]
\end{definition}

Since the above definition is symmetric for $X$ and $Y$,
it follows that $\VZ \vdash X + Y = Y + X$.
It remains to show the associativity for this function.

\begin{lemma}
\label{t:associative}
$\VZbar\vdash X+(Y+Z) = (X+Y)+Z$.
\end{lemma}

\begin{proof}
It suffices to show that for $i < |X| + |Y| + |Z|$,
\[(X+(Y+Z))(i)\lra((X+Y)+Z)(i).\]
This is equivalent to
\begin{equation*}
X(i)\oplus(Y+Z)(i)\oplus\varphi_+(i,X,Y+Z)
\lra
(X+Y)(i)\oplus Z(i)\oplus\varphi_+(i,X+Y,Z).
\end{equation*}
From \eqref{e:carry}, the above is simplified to
\[\varphi_+(i,Y,Z)\oplus\varphi_+(i,X,Y+Z)
\lra
\varphi_+(i,X,Y)\oplus\varphi_+(i,X+Y,Z).\]

Let $a_i$, $b_i$, $c_i$ and $d_i$ denote
$\varphi_+(i,Y,Z)$, $\varphi_+(i,X,Y+Z)$, $\varphi_+(i,X,Y)$ and
$\varphi_+(i,X+Y,Z)$ respectively.
We need to show that
\[a_i \oplus b_i \lra c_i \oplus d_i,\]
for $i < |X| + |Y| + |Z|$.
We will prove a stronger result,
i.e., $(a_i, b_i)$ is a permutation of $(c_i, d_i)$.
In particular, we will show by induction on $i$ that
\begin{equation}
\label{e:associative}
(a_i\wedge b_i\lra c_i\wedge d_i)
\wedge
(a_i\vee b_i\lra c_i\vee d_i).
\end{equation}

The base case is trivial, since
\[\VZ\vdash\neg a_0\wedge\neg b_0\wedge \neg c_0\wedge \neg d_0.\]
The induction step follows from the inductive evaluation of $a_{i}$
$b_i$, $c_i$ and $d_{i}$:
\begin{eqnarray*}
a_{i+1} & = & [Y(i)\wedge Z(i)]\vee[(Y(i)\oplus Z(i))\wedge a_i],\\
b_{i+1} & = & [X(i)\wedge(Y(i)\oplus Z(i)\oplus a_i)]\vee
[(X(i)\oplus Y(i)\oplus Z(i)\oplus a_i)\wedge b_i],\\
c_{i+1} & = & [X(i)\wedge Y(i)]\vee[(X(i)\oplus Y(i))\wedge c_i],\\
d_{i+1} & = & [Z(i)\wedge(X(i)\oplus Y(i)\oplus c_i)]\vee
[(X(i)\oplus Y(i)\oplus Z(i)\oplus c_i)\wedge d_i].
\end{eqnarray*}
It remains to verify \eqref{e:associative} for $i + 1$.
This can be done by using the induction hypothesis and the above properties,
together with checking all possible values of $X(i), Y(i)$ and $Z(i)$.
Details are omitted.
\end{proof}

\section{Proving the Pigeonhole Principle in \VTCZ}
\label{s:PHP}

We give an example of reasoning in \VTCZ\ by formalizing and
proving the Pigeonhole Principle (\PHP) in \VTCZ.
This principle states that for any mapping from a set of $a$ numbers to
a set of $(a - 1)$ numbers, there must be 2 numbers in the domain
that have the same image.
We will formalize and prove this principle in \VTCZ.
In the following definition, the mapping is represented by the set $X$ of
pairs of pre-images and images
($X(y, z)$ holds if $y$ is the image of $z$).
\begin{theorem}
\label{t:PHP}
$$
\VTCZ \vdash \forall z\le a\ \exists y<a\ X(y,z)\ \supset\
\exists y<a\ \exists z_1\le a\ \exists z_2<z_1,\ X(y, z_1)\wedge X(y, z_2).
$$
\end{theorem}

Proving \PHP\ involves formalizing a number
of concepts, such as set union, total number of bits in an array, etc.
We will define these functions below, and it is straightforward that
they are members of \LFTCZ.

\noindent
{\bf Union}:
$\Union(b, X, Y)(z) \lra z<b\wedge(X(z)\vee Y(z))$.

We interpret $Z$ as an array of $a$ rows, and each row has length bounded by $b$.

\noindent
{\bf Finite union}:
$\FiniteUnion(a, b, Z)(z) \lra z<b\wedge\exists y<a Z^{[y]}(z)$.

\noindent
{\bf Total number of bits in an array}:
$\totNumones(a, b, Z) = \numones(ab, F_0(a, b, Z))$, where
$F_0(a, b, Z)$ is the function of $Z$ which concatenates all the rows of the array $Z$:
$$F_0(a, b, Z)(ax + y) \lra Z^{[x]}(y),\ \text{for $x < a$, $y < b$}$$

\begin{lemma}
\label{t:PHP-aux}
The following are theorems of \VTCZbar:\\
{\em a)} $\numones(b, \Union(b,X,Y))\le \numones(b, X)+\numones(b, Y)$.\\
{\em b)} $\totNumones(a+1,b,Z) = \totNumones(a,b,Z) + \numones(b, Z^{[a]})$.\\
{\em c)} $\numones(b, \FiniteUnion(a,b,Z))\le\totNumones(a,b,Z)$.\\
{\em d)} $\forall x<a\ \numones(b, Z^{[x]}) \le k\ \supset\ \totNumones(a,b,Z) \le ak$.
\end{lemma}

\begin{proof}
Part a) is proved by induction on $b$. 
Part b) is proved by noting that $F_0(a+1, b, Z)$ is the concatenation of
$F_0(a, b, Z)$ and $Z^{[a]}$.

For c), the proof is by induction on $a$.
The base case is straightforward.
For the induction step, note that
$$\FiniteUnion(a+1,b,Z) = \Union(b,\FiniteUnion(a,b,Z),Z^{[a]})$$
We have
\[
\begin{array}{cl}
& \numones(b, \FiniteUnion(a+1,b,Z)) \\
 = & \numones(b, \Union(b, \FiniteUnion(a,b,Z),Z^{[a]}))\\
 \le & \numones(b, \FiniteUnion(a,b,Z)) + \numones(b, Z^{[a]})\ \ \text{(by a.)} \\
 \le & \totNumones(a,b,Z) + \numones(b, Z^{[a]})\ \ \text{(by the I.H.)} \\
 = & \totNumones(a+1,b,Z)\ \ \text{(by b.)}
\end{array}
\]

Finally, part d) is proved by induction on $a$, using part b).
\end{proof}

\begin{proof}
[Proof of Theorem~\ref{t:PHP}]
We have to show that there exists a row of $X$ that contains at least 2 bits.
We will prove by contradiction, by showing that if every row
of $X$ has at most 1 bit, then the total number of bits in the
array $X$ is at most $a$.
On the other hand, the union of the rows of $X$ has $(a+1)$ bits.
This contradicts part c) of Lemma~\ref{t:PHP-aux}.
Details are as follows.

Suppose that $\forall y<a\ \numones(a+1, X^{[y]})\le 1$.
Then part d) of Lemma~\ref{t:PHP-aux} implies $\totNumones(a, a+1, Z)\le a$.
By part c) of Lemma~\ref{t:PHP-aux}, $\numones(a+1, \FiniteUnion(a,a+1,Z)) \le a$.
However, it is obvious that $\forall z\le a\ \FiniteUnion(a,a+1,Z)(z)$.
By a simple
induction argument, this implies $\numones(a+1, \FiniteUnion(a,a+1,Z)) = a+1$,
a contradiction.
\end{proof}

\end{document}